\documentclass[10pt,a4paper]{article}
\usepackage{jcappub}

\pdfoutput=1

\usepackage{%
graphicx,%
array,%
bm,%
booktabs,%
multirow%
}
\usepackage[utf8]{inputenc}
\usepackage[english]{babel}
\usepackage{amsmath,amsfonts,amssymb}
\usepackage[usenames,dvipsnames,svgnames]{xcolor}
\usepackage[T1]{fontenc}
\usepackage{capt-of}
\newcommand{\be}{\begin{equation}}
\newcommand{\ee}{\end{equation}}

\newcommand{\mytablen}{.7em}

\renewcommand{\tilde}{\widetilde}
\def\FT{F(\mathcal T)}

\title{Model-independent limits and constraints on extended theories of gravity from cosmic reconstruction techniques}
\author[a,b]{\'Alvaro de la Cruz-Dombriz}
\author[a,b,c]{Peter K. S. Dunsby}
\author[a,b,d,e]{Orlando Luongo}
\author[a,b]{Lorenzo Reverberi}

\affiliation[a]{Department of Mathematics and Applied Mathematics, University of Cape Town, Rondebosch 7701, Cape Town, South Africa.}
\affiliation[b]{Astrophysics, Cosmology and Gravity Centre (ACGC), University of Cape Town, Rondebosch 7701, Cape Town, South Africa.}
\affiliation[c]{South African Astronomical Observatory,  Observatory 7925, Cape Town, South Africa}
\affiliation[d]{Dipartimento di Fisica, Universit\`a di Napoli ''Federico II'', Via Cinthia, I-80126, Napoli, Italy.}
\affiliation[e]{Istituto Nazionale di Fisica Nucleare (INFN), Via Cinthia, I-80126, Napoli, Italy.}

\emailAdd{alvaro.delacruzdombriz@uct.ac.za}
\emailAdd{peter.dunsby@uct.ac.za}
\emailAdd{luongo@na.infn.it}
\emailAdd{lorenzo.reverberi@uct.ac.za}

\arxivnumber{1608.03746}

\abstract{
The onset of dark energy domination depends on the particular gravitational theory driving the cosmic evolution. Model independent techniques are crucial to test the both the present $\Lambda$CDM cosmological paradigm and alternative theories, making the least possible number of assumptions about the Universe. In this paper we investigate whether cosmography is able to distinguish between different gravitational theories, by determining bounds on model parameters for three different extensions of General Relativity, namely quintessence, $\FT$ and $f(R)$ gravitational theories. We expand each class of theories in powers of redshift $z$ around the present time, making no additional assumptions. This procedure is an extension of previous work and can be seen as the most general approach for testing extended theories of gravity through the use of cosmography. In the case of $\FT$ and $f(R)$ theories, we show that some assumptions on model parameters often made in previous works are superfluous or even unjustified.

We use data from the Union 2.1 supernovae catalogue, baryonic acoustic oscillation data 
and $H(z)$ differential age compilations, which probe cosmology on different scales of the cosmological evolution. We perform a Monte Carlo analysis using a Metropolis-Hastings algorithm with a Gelman-Rubin convergence criterion, reporting 1--$\sigma$ and 2--$\sigma$ confidence levels. To do so, we perform two distinct fits, assuming only data within $z<1$ first and then without limitations in redshift. We obtain the corresponding numerical intervals in which coefficients span, and find that the data is compatible the $\Lambda$CDM limit of all three theories at the 1-$\sigma$ level, while still compatible with quite a large portion of parameter space. We compare our results to the truncated $\Lambda$CDM paradigm, demonstrating that our bounds divert from the expectations of previous works, showing that the permitted regions of coefficients are significantly modified and in general widened with respect to values usually reported in the existing literature.
Finally, we test the extended theories through the Bayesian selection criteria AIC and BIC.
}

\begin{document}

\maketitle
\flushbottom

\section{Introduction}\label{S1:Introduction}

One of the most important challenges in modern cosmology is to determine whether the theory of General Relativity (GR) is
the best gravitational theory able to describe the dynamics of the Universe on all scales. In particular, since the discovery of the cosmic acceleration~\cite{Riess:1998cb,Perlmutter:1997zf,Schmidt:1998ys,Perlmutter:1998np,Riess:2004nr} and in view of recent developments related to the discovery of gravitational waves~\cite{nextnobel}, determining a completely self-consistent description of the gravitational action has become even more essential~\cite{review1}. Indeed, on the one hand observations of supernovae suggest that our Universe is currently speeding up, while on the other hand structure formation constraints and causal requirements on the cosmic angular distribution suggest the presence of three unknown ingredients, namely dark energy, dark matter and the inflaton field~\cite{review2}. Thus, in lieu of invoking new ingredients in Einstein's equations, extensions of GR turn out to be a very natural landscape to investigate where Einstein's gravity might break down~\cite{Capozziello:2011et}. Furthermore, some extended theories of gravity have the advantage of being able to describe both the current cosmic evolution and the whole expansion history of the Universe at higher energy scales~\cite{ReviewsfR3}.

However, the main disadvantages of any modification of Einstein's gravity relate to the physical motivation of these theories and any instabilities that emerge as a result of introducing extra terms. In particular, among all plausible modifications, several approaches successfully reproduce the cosmic evolution with the same accuracy as the Concordance cosmological  $\Lambda$CDM model, leading to a degeneracy problem~\cite{deg:1,deg:2,deg:3,deg:4}.
A possible way of alleviating such a degeneracy is to combine different measurements with the aim of reducing the phase space of free parameters.
Nevertheless, the caveat of any measurement is that one is often forced to assume the statistical validity of a given cosmological model a priori. In such a way, measures and error propagation lead to the particular framework under examination, without being conclusive in selecting models. For those reasons, amongst several statistical treatments, the class of model-independent techniques becomes essential to guarantee that the statistical outcomes do not depend upon the choice of the model itself. Research of the so-called \emph{cosmography} of the Universe belongs to this class and has drawn much attention in recent years~\cite{Weinberg:Gravitation_and_Cosmology,cosmo1:2,cosmo2,cosmo3}. Cosmography is essentially a Taylor expansions in cosmology. Derivatives of the scale factor $a$ (or of the Hubble parameter $H\equiv \dot a/a$) are naturally model-independent, but they are strongly related to the values of the parameters of any extended model. Sometimes, one refers to cosmography as \emph{cosmo-kinetics}, in which one expands the main observable quantities in terms of the cosmic time.\footnote{Or alternatively in terms of the redshift $z$.}~\cite{cosmo5:1,cosmo5:2}

The main purpose of this work is to extend previous literature on cosmographic approaches in the field of extended theories ({\it c.f.}~\cite{Aviles:2012ir,bc,cosmo9} and references therein). In doing so, we shall match cosmography with different classes of extended theories of gravity, making no further assumptions in the parameter space of the gravitational theories under scrutiny. This was in fact a common denominator shared by all previous works, which either prevented capture of essential trends in the classes of models under consideration or over/underestimated the power of cosmography as a competitive technique capable of reconstructing or constraining underlying theories. We shall consider different tests such as type Ia supernovae Union 2.1 compilation, baryonic acoustic oscillation (BAO) data and Hubble rate measurements.

In the following sections we shall deal with competitive extended theories of gravity, namely quintessence theories, extended teleparallel gravity theories, dubbed $F(\mathcal{T})$, and $f(R)$ gravity theories.

Cosmographic studies in these theories have been limited to recent research~\cite{cosmo9, delaCruz-Dombriz:2016rxm}
which proved how limited such a technique is when compared to Gaussian processes of reconstruction~\cite{Nair:2013sna}. Moreover the usual approximation that the matter content today can be related to the cosmographic parameter $q_0$ as happens in $\Lambda$CDM was made, limiting the generality of the obtained results therein.

The second class of theories in our analysis is the so-called extended Teleparallel gravity theories, dubbed $F(\mathcal{T})$. In these theories the covariant action is written in terms of an arbitrary function of the torsion scalar $\mathcal{T}$, which indeed extends the
usual teleparallel gravity ({\it c.f.}~\cite{Hehl:1976kj,Hayashi:1979qx,Wilczek:1998ea,Obukhov:1983mm,Benn:1980ea,Aldrovandi:2013wha} for further insight), a gravitational theory associated to the translation group where a Minkowskian tangent spacetime is associated with every point of the spacetime. Thus, these theories are constructed, instead of the usual Levi-Civita connection, in terms of the Weitzenb\"ock connection, which induces a non-zero torsion but a vanishing Riemann tensor. This fact enables us to transport the so-called vierbeins/tetrads fields in parallel, providing the name of the theory.

 Dark energy can be accommodated in the framework of these theories~\cite{Bengochea:2008gz} and since, unlike $f(R)$ theories, the field equations are still second-order, gravitational waves do not exhibit extra modes~\cite{Bamba:2013ooa}. On the other hand, as it is widely known, extensions of Teleparallel gravity are not Lorentz invariant (see ref.~\cite{Li:2010cg,Sotiriou:2010mv}), and suffer from issues with acausality and non-uniqueness~\cite{Ong:2013qja, Izumi:2013dca, Chen:2014qtl}. The field equations will be sensitive to the choice of tetrads and consequently, the determination of the correct tetrad fields, leading to a metric tensor with some desirable symmetries, has attracted some attention in recent years. For instance studies have been devoted to cosmological solutions~\cite{Ferraro:2011us}, the Schwarzschild solution~\cite{Bohmer:2011si,Meng:2011ne,Dong:2012en,Iorio:2012cm,Wang:2011xf,HamaniDaouda:2011iy,Daouda:2011rt,Ferraro:2011ks}, the validity of Birkhoff's theorem~\cite{Tamanini:2012hg, Meng:2011ne, Boehmer:2011gw} and finally junction conditions~\cite{delaCruz-Dombriz:2014zaa}, proving the interest in such theories.
Nonetheless, very few references have been devoted so far to cosmographic techniques for this class of theories. In~\cite{Capozziello:2015rda}, the authors extracted some constraints on the redshift transition determining the onset of cosmic acceleration, employing cosmography to obtain bounds on the viable $F(\mathcal{T})$ forms and performed a Monte Carlo fitting using supernovae data.

Finally we shall complete our analysis with the study of theories including invariants constructed with the Riemann tensor contractions. Pioneering works~\cite{Gurovich:1979xg,Starobinsky:1979ty,Starobinsky:1980te} found that one-loop quantum corrections to the vacuum expectation value of the energy-momentum tensor generate terms containing higher-order curvature invariants, such as $R^2$, $R_{\mu\nu} R^{\mu\nu}$, etc., with typical couplings of the order of the Planck mass by the appropriate (negative) power. Therefore, such quantum corrections are relevant only at very large curvatures which originally gave rise to the interest in $f(R)$ theories, which might be thought of as the only local, metric-based and generally coordinate invariant and stable modifications of gravity~\cite{Woodard:2006nt, Biswas:2011ar}.
This perspective has now changed. Recent results also show that when quantum corrections or string theory are taken into account, higher-order curvature invariants may appear naturally in the effective low-energy Lagrangian~\cite{Birrell:1982ix,Buchbinder:1992rb,Vilkovisky:1992pb}. Anyhow, $f(R)$ theories have remained a paradigmatic example of extended theories of gravity ({\it c.f.}~\cite{ReviewsfR1,ReviewsfR2,ReviewsfR3,ReviewsfR4,Capozziello:2011et,ReviewsfR6,ReviewsfR7,ReviewsfR8} for extensive reviews and citations therein) capable of avoiding several classes of instabilities, providing inflationary mechanisms~\cite{Starobinsky:1980te,applicationsstaro1, applicationsstaro2, applicationsstaro3, applicationsstaro4, Bamba:2014daa, Bamba:2014wda, Dombrizinflation}, accounting for the dark matter component~\cite{Cembranos:2008gj, Capozziello:2012ie}, ensuring the correct growth of large-scale structures~\cite{perturbations1,perturbations2,perturbations3,perturbations4} and passing a multitude of  tests in several astrophysical scales~\cite{tests1,tests2,tests3,tests4,tests5,tests6,delaCruz-Dombriz:2015tye,tests8}. Some cosmographic studies were made in the context of $f(R)$ theories~\cite{Capozziello:2014zda, Aviles:2012ir} although the present cosmological values of the first and second derivatives of the gravitational Lagrangian were therein fixed to their GR counterparts.\footnote{Bear in mind that cosmological values today, for such derivatives, may differ from their GR counterparts and still produce viable cosmological models.} However, whenever  ${\rm d}^2f/{\rm d}R^2=0$, either a singularity or instability occurs~\cite{Pogosian:2007sw,delaCruz-Dombriz:2015tye}. In the event of those parameters not being fixed, a one-to-one correspondence between the $f(R)$-derivatives and the cosmographic parameters is no longer possible~\cite{cosmo9, delaCruz-Dombriz:2016rxm} which forces a sensible prior assumption over these derivatives or complementary tests. Anyhow, these attempts led to limited constraints on $f(R)$ models~\cite{cosmo9, delaCruz-Dombriz:2016rxm} that we intend to extend herein.

The paper is organised as follows:\footnote{Unless otherwise specified, natural units $\hbar=k_B=c=1$ will be used throughout this paper and $8\pi G\equiv8\pi G$ with $G$ being the standard gravitational constant.}
In Section \ref{S2:Cosmography} we sketch the approach to generalise the cosmographic method to extended theories of gravity. In \ref{S2:Subsection1: Rudiments}, we provide a quick review of the cosmographic method and how both luminosity and angular distances can be parametrised using cosmographic parameters.
Then in \ref{S2:Subsection2: Cosmography beyond LCDM} we describe our approach and a number of limitations that the cosmographic method suffers from when it is applied to higher-order theories of gravity, for example the need to introduce extra parameters and priors or physical intervals on these parameters. We also presented the catalogues used in this paper,  as well as the expected usefulness of each dataset.
In Section \ref{S3:Constraining} we present the required dictionaries connecting the cosmographic parameters and the appropriate parameters which characterise the theories under consideration, namely quintessence, $F(\mathcal{T})$ and $f(R)$ theories in \ref{k-essence theories}, \ref{f(T) theories} and \ref{f(R) theories} respectively.
Then in Section \ref{S4a:results} we present our main results, after extensive use of Markov chain Monte Carlo (MCMC) analyses for each of the aforementioned theories, providing the used priors - if any - and the statistical constraints for the parameters of each theory. This enables us to reconstruct the best fits. Combined and independent analyses of Supernovae, BAO and $H(z)$ data are then presented.
We end the paper in Section \ref{S4:Conclusions}, giving our conclusions and discussing future strategies.

\section{Cosmographic approach in extended theories}
\label{S2:Cosmography}

In this section, we shall briefly present the role of cosmography in standard cosmology, with particular attention devoted to its application to extended theories of gravity to be developed in the upcoming sections.

\subsection{Basics of Cosmography}
\label{S2:Subsection1: Rudiments}

In its modern interpretation, cosmography makes sole use of the Cosmological Copernican principle, without any further assumptions~\cite{cosmo6}, which naturally leads to the Friedmann-Lema\^{i}tre-Robertson-Walker (FLRW) metric describing a homogeneous and isotropic Universe. We can expand the cosmological scale factor $a(t)$ around the present time $t_0$ as:
\begin{equation}\label{adit}
a(t)=\sum_{j=0}\frac{a^{(j)}(t_0)}{j!}\,(t-t_0)^j\,,
\end{equation}
and one can easily show that its derivatives are intimately related to observable quantities of cosmological interest.

In particular, the cosmographic coefficients:\footnote{Usually defined as \emph{cosmographic series}.}, are determined as scale factor derivatives evaluated at the present time~\cite{cosmo7}
\begin{equation}
\label{pinza}
H=\frac{\dot{a}}{a}\,,\quad q=-\frac{\ddot{a}}{aH^2}\,,\quad j=\frac{a^{(3)}}{aH^3}\,,\quad s=\frac{a^{(4)}}{aH^4}\,,
\end{equation}
where dots represent cosmic time derivatives. Also, in terms of the observable Hubble rate $H$ the cosmographic coefficients become:
\be\label{eq:CSoftime}
q=-\frac{\dot{H}}{H^2} -1\,,\quad j=\frac{\ddot{H}}{H^3}-3q-2\,,\quad s=\frac{H^{(3)}}{H^4}+4j+3q\left(q+4\right)+6\,,
\ee
Cosmography may quantify the amount and time evolution of dark energy which is effectively required to permit the Universe to accelerate, as indicated by current observations.

\subsection{Cosmography Beyond the Standard Cosmological Model}
\label{S2:Subsection2: Cosmography beyond LCDM}

As mentioned above, we intend to consider three extensions of GR, namely quintessence, $\FT$ and $f(R)$ theories. The aim is to employ the cosmographic approach, that is to perform a Taylor expansion in powers of time or redshift around the present time, in order to derive constraints on model parameters.

One should always be careful when combining modified gravity theories and cosmological observations. Indeed, cosmography itself relies on the assumptions of homogeneity and isotropy, so clearly it would be impossible to test any modified gravity theory or cosmology without such properties. As far as SNIa and $H(z)$ measurements are concerned, they only depend on the expansion history so we can safely use such data to test modified gravity theories. BAO measurements present subtleties (see also below), because the BAO scale can be used as a standard ruler only under certain assumptions about the evolution of (linear) perturbations, which in general differs between $\Lambda$CDM and alternative theories. However, it has been shown~\cite{Bellini:2015oua} for a wide class of theories, which include quintessence and $f(R)$, that one can indeed use BAO as a test of gravity. For $\FT$ theories, the situation is less clear because of several problems affecting these theories as was mentioned in the introduction. Nevertheless, a comprehensive analysis of BAO within $\FT$ theories is well beyond the scope of this work, and hardly compatible with the notion of generality we would like to preserve. Implicitly, our results on $\FT$ (and indeed other models as well) assume that we are dealing with reasonably ``well-behaved'' theories.

We run multiple MCMC reconstruction chains using a Metropolis-Hastings algorithm~\cite{metro:1,metro:2,metro:3}, where we have previously defined convenient combinations of the derivatives in the gravitational Lagrangians for every class of theories under investigation. To do so, we consider derivatives of $V(\phi)$, $\FT$ and $f(R)$ with respect to the redshift $z$ and we compare with Union2.1~\cite{Amanullah:2010vv}, BAO~\cite{Beutler:2011hx, Ross:2014qpa, Anderson:2013zyy, Padmanabhan:2012hf} and $H(z)$~\cite{Farooq:2012ju, Farooq:2013hq} datasets.

Before proceeding, let us start by expressing both
the luminosity ($d_L$) and angular ($d_A$) distances in the cosmographic  expansions as follows:
\begin{subequations}\label{distanzadilumin1e2}
\begin{align}
   d_{L,A}(z)  &= \frac{z}{H_0}(1+\eta_1z+\eta_2z^2+\eta_3z^3+\ldots)\,,\\
   \eta_{1,L}  &= \frac{1}{2} - \frac{q_0}{2}\,,\\
    \eta_{2,L}  &= -\frac{1}{6} -\frac{j_0}{6} + \frac{q_0}{6} + \frac{q_0^2}{2}\,,\\
    \eta_{3,L}  &= \frac{1}{12} + \frac{5 j_0}{24} - \frac{q_0}{12} + \frac{5 j_0 q_0}{12}-\frac{5 q_0^2}{8} - \frac{5 q_0^3}{8} + \frac{s_0}{24}\,\\
    \eta_{1,A} &=-\Bigl(\frac{3}{2} + \frac{q_0}{2} \Bigr)\,,\\
    \eta_{2,A} &=\frac{11}{6} -\frac{j_0}{6} + \frac{7 q_0}{6} + \frac{q_0^2}{2}\,,\\
    \eta_{3,A} &=-\frac{25}{12} + \frac{13 j_0}{24} - \frac{23 q_0}{12} + \frac{5 j_0 q_0}{12}
    -  \frac{13 q_0^2}{8} - \frac{5 q_0^3}{8} + \frac{s_0}{24}\,.
\end{align}
\end{subequations}

The above coefficients, entering eq. (\ref{distanzadilumin1e2}),  as evaluated in the current epoch, are called the deceleration parameter, $q_0$, which specifies whether the Universe is experiencing either an accelerating ($-1<q_0<0$) or decelerating ($q_0>0$) phase; the jerk term, $j_0$, which gives us information about the change of acceleration; and the
the snap parameter $s_0$, which defines the slope of the luminosity curve at higher redshifts. For example, at the level of $\Lambda$CDM model with $\Omega_m=0.318$, $\Omega_k=0$, one gets $q_0=-0.523$, $j_0=1$ and $s_0=-0.431$.

\def\mn{{\mu\nu}}
\def\be{\begin{equation}}
\def\ee{\end{equation}}

\subsection{Problems and Shortcomings of Cosmography}
\label{sec:problems}

One of the shortcomings of the cosmographic approach appears obvious when one considers that the expansion parameter (the redshift) is not necessarily small, in fact it can assume values larger than unity. This can generate convergence problems, and makes the truncation at a finite expansion order at least questionable. To overcome this problem, cosmographic reconstructions often employ auxiliary parametrisations of cosmic distances, which commonly involve new choices of independent variables\footnote{Two immediate examples are $y_1=\frac{z}{1+z}$ and $y_2=\arctan(z)$. For further examples see for instance~\cite{cosmo8}.} built up in terms of the redshift $z$. However, the viability of such analyses has recently been put into question, due to increasing propagation of errors and difficulties in understanding what the most suitable parametrisation at a statistical level is~\cite{cosmo9,delaCruz-Dombriz:2016rxm}. Such combinations of redshift $z$ seemed to lead to biased results, not being for instance even able to unveil $\Lambda$CDM as the theory responsible for mock data precisely generated from an exact $\Lambda$CDM model. Thus we have opted to perform the cosmographic expansions in terms of redshift $z$.

We mention that we have carried out our whole analysis in terms of both $y_1$ and $y_2$ as well, although we do not present the results in this paper. In both cases, the mean values are compatible with those of the $z$ analysis, but the standard errors are much larger, up to several orders of magnitude in the case of $y_1$. This somewhat confirms the problems with alternative variables mentioned above.

In this work, we are mostly interested in testing the potential of the cosmographic approach to constrain modified gravity theories. In other words, what is interesting for us the \textit{precision} of the results, simply put the ``error bars'', rather than their \textit{accuracy}, i.e. the statistical bias. As we will see below, one can easily recognize such bias in how results for the (truncated) $\Lambda$CDM model are modified when considering $z<1$ and all--$z$ datasets. However, error bars remain of the same order or magnitude and actually practically identical in the two cases.

Unsurprisingly (see below), our results our compatible with the $\Lambda$CDM limit of each model within 1-$\sigma$. Hence, regardless of the statistical bias introduced by using a finite expansion despite $z>1$, it is ultimately the precision of the results what has the biggest impact on our potential to constrain alternative gravity models. The goal of this paper is not to show how \textit{much} cosmography is able to tell us about modified gravity theories, but rather how \textit{little}. This is the reason why we have decided to use the full dataset, including data points at $z>1$.

\section{Constraining Alternative Theories with Cosmography}
\label{S3:Constraining}

In this section, we consider the above requirements of cosmography to relate Universe's kinematics to particular classes of dark energy models. In particular, cosmography turns out to be a model independent treatment to obtain bounds on cosmic observables, but by virtue of~\eqref{eq:CSoftime}, one can match the cosmographic expectations with the theoretical predictions of any dark energy model. First of all, we must find a way to uniquely express the cosmographic parameters in terms of model parameters, so that constrains on the former can be translated into constraints on the latter, which are what we are most interested in.

\subsection{Quintessence Theories}
\label{k-essence theories}

Let us consider as a first example of our method a minimally coupled scalar field $\phi$ plus the standard Einstein-Hilbert action, i.e., \textit{quintessence} dark-energy models~\cite{Fujii:1982ms,Ratra:1987rm,Caldwell:1997ii,Wetterich:1987fm} of the form:
\begin{equation}
\mathcal{S}_\phi\,=\,\int {\rm d}^{4}x\sqrt{-g}\left[ \frac{R}{16\pi G}
 - \frac{1}{2} g^\mn \partial_\mu \phi\,\partial_\nu\phi -V(\phi )+\mathcal{L}_{m}\right]\ ,
\label{ST1_orl}
\end{equation}
where  $\mathcal{L}_m$ is the matter Lagrangian density and $V(\phi)$ is the scalar field potential. In a FLRW Universe, the field equations read
\begin{subequations}
\label{eq_scalar_tensor_field_eqs}
\begin{align}
& H^{2} = \frac{8\pi G}{3}\left[\rho _{m} + \frac{\dot\phi^2}{2} + V(\phi)\right]\,, \label{eq_scalar_tensor_Friedmann}\\
& \dot{H} = -\frac{8\pi G}{2}\left[ \rho _{m}(1+w_m) + \dot \phi^2 \right]\,, \label{eq_scalar_tensor_Raycha}\\
& \ddot{\phi}+3H\dot{\phi} + V_\phi =0\,,
\label{eq_scalar_tensor_scalar}
\end{align}
\end{subequations}
where we have assumed that matter is well-described by a perfect fluid with constant equation of state $P_m=w_m\rho_m$. Indeed, we will consider the dust case $w_m=0$, which is an excellent approximation for both baryonic and dark matter at late times ($z\lesssim 1.5$). Thus, we will parametrise the matter energy density in the usual way:
\be\label{eq_definition_Omega_m}
8\pi G\rho_m = 3H_0^2\Omega_m(1+z)^3\,.
\ee
Let us also define:
\begin{subequations}\label{eq_definitions_ST}
\begin{align}
& \tilde V_0 = \frac{8\pi G}{H_0^2}V_0 - 3(1-\Omega_m)\,,\\
& \tilde V_i = \frac{8\pi G}{H_0^2}\left.\frac{\partial^i V}{\partial z^i}\right|_{z=0}\,\,,\,\, i=1,2\,,\\
\end{align}
\end{subequations}
This choice guarantees that these parameters are dimensionless and that $\Lambda$CDM corresponds to the limit in which all parameters vanish, i.e. $\tilde V_i = 0$ $(i=0,1,2)$. Then, using all the expressions in eqs.~(\ref{eq_scalar_tensor_field_eqs}), the first and second derivatives of~(\ref{eq_scalar_tensor_Raycha}) and the first derivative of~(\ref{eq_scalar_tensor_scalar}), we are able to express $q,j,s$ in terms of the three parameters $\tilde V_0, \tilde V_1, \tilde V_2$ plus $\Omega_m$. Thus the cosmographic parameters when evaluated today read
\begin{subequations}
\label{eq_cosmography_ST}
\begin{align}
 q_0 &= -1 + \frac{3\Omega_m}{2} - \tilde V_0\,,\\
 j_0 &= 1 - 3\tilde V_0 - \tilde V_1\,,\\
 s_0 &= 1 - \frac{9\Omega_m}{2} - 3\tilde V_0^2 + \tilde V_0\left(18-\tilde V_1 + \frac{9\Omega_m}{2}\right)  
+ \frac{3\tilde V_1(2+\Omega_m)}{2} + \tilde V_2\,.
\end{align}
\end{subequations}

\subsection{\texorpdfstring{$F(\mathcal{T})$}{F(T)} Theories}
\label{f(T) theories}

Let us now consider extended theories of teleparallel gravity~\cite{Ferraro:2006jd, Bengochea:2008gz, Linder:2010py, Myrzakulov:2010vz}, whose action reads
\begin{equation}
 {\mathcal S}= \int {\rm d}^4x \, e \left[
\frac{\FT}{2{\kappa}^2}+{\cal L}_{m}
\right]\;,
\label{actionfT}
\end{equation}
where $e= \det \left(e^A_\mu \right)=\sqrt{-g}$ and $\mathcal T$ is the torsion scalar (see e.g.~\cite{Aldrovandi:2013wha}). The field equations for a FLRW Universe read
\begin{subequations}\label{eq_fT_field_eqs}
\begin{align}
& H^2 = \frac{\rho_m + \rho_{\FT}}{3}\,, \label{eq_fT_Fried}\\
& \dot H = -\frac{\rho_m + \rho_{\FT} + P_{\FT}}{3}\,, \label{eq_fT_Raycha}
\end{align}
\end{subequations}
with
\be
\rho_{\FT} = \frac{J_1}{2}\,,\quad P_{\FT} = -\frac{J_1 + 4 J_2}{2}\,,
\ee
and
\be
J_1 = -\mathcal T - F + 2\mathcal T F'\,,\quad J_2 = \dot H(1-F' - 2\mathcal T F'')\,.
\ee
with prime above denoting derivative with respect to $\mathcal T$. Moreover, the torsion scalar is simply $\mathcal T = -6H^2$. Let us define:
\begin{subequations}\label{eq_definitions_fT}
 \begin{align}
  F_i = \mathcal T^{i-1}\left.\frac{\partial^i F}{\partial \mathcal T^i}\right|_{z=0}\,,
 \end{align}
\end{subequations}
which are dimensionless quantities. Then, using~(\ref{eq_fT_field_eqs}) and the first redshift derivative of~(\ref{eq_fT_Raycha}) yields
\begin{subequations}\label{eq_cosmography_fT_initial}
 \begin{align}
  q_0 &= -1 + \frac{3\Omega_m}{2(F_1 + 2F_2)}\,,\\
  j_0 &= 1 - \frac{9\Omega_m^2(3F_2+2F_3)}{2(F_1 + 2F_2)^3}\,,\\
  s_0 &= 1 + \frac{-9\Omega_m}{2(F_1 + 2F_2)} + \frac{45(3F_2 + 2F_3)\Omega_m^2}{2(F_1 + 2F_2)^3}  
 + \frac{ 27(3F_2 + 12F_3 + 4F_4)\Omega_m^3}{4(F_1 + 2F_2)^4}\,+  \notag \\
  &\qquad
  + \frac{-81(3F_2 + 2F_3)^2\Omega_m^3}{2(F_1 + 2F_2)^5}\,.
 \end{align}
\end{subequations}
We have also used the usual parametrisation~(\ref{eq_definition_Omega_m}). It appears that we are left with four independent parameters (besides $\Omega_m$), namely $F_i$ ($i=1,2,3,4$), which allow us to only obtain three cosmographic parameters $(q,j,s)$. However, rescaling all quantities as follows:
\be
\label{eq_definitions_fT_FINAL}
\tilde\Omega_m = \frac{\Omega_m}{F_1}\,,\qquad
 \tilde F_j = \frac{F_j}{F_1}\quad (j = 2,3,4)\,,
\ee
we are able to eliminate $F_1$ from~(\ref{eq_cosmography_fT_initial}) entirely, finding
\begin{subequations}\label{eq_cosmography_fT_FINAL}
 \begin{align}
 q_0 &= -1 + \frac{3\tilde\Omega_m}{2(1 + 2\tilde F_2)}\,,\\
  j_0 &= 1 - \frac{9\tilde\Omega_m^2(3\tilde F_2+2\tilde F_3)}{2(1 + 2\tilde F_2)^3}\,,\\
  s_0 &= 1 + \frac{-9\tilde\Omega_m}{2(1 + 2\tilde F_2)} + \frac{45\left(3\tilde F_2 + 2\tilde F_3\right)\tilde\Omega_m^2}{2(1 + 2\tilde F_2)^3}  
  + \frac{ 27(3\tilde F_2 + 12\tilde F_3 + 4\tilde F_4)\tilde\Omega_m^3}{4(1 + 2\tilde F_2)^4}  \,+
  \notag \\&\qquad
  + \frac{-81(3\tilde F_2 + 2\tilde F_3)^2\tilde\Omega_m^3}{2(1 + 2\tilde F_2)^5}\,.
\end{align}
\end{subequations}
As in the case of quintessence theories, the parametrisation above is such that $\Lambda$CDM is recovered when all the introduced parameters are zero, i.e., $\tilde F_i = 0$ $(i=2,3,4)$. Furthermore, the expression for the present value of the function $\FT$ yields:
\be
\tilde F_0 = \left.\frac{\FT}{\mathcal T}\right|_{z=0} = 2 - \tilde\Omega_m\,.
\label{F_0}
\ee
Unlike $f(R)$ theories (see below), $\tilde F_0$ in eq.~(\ref{F_0}) is determined from the field equations with no dependence on parameters other than $\Omega_m$, which is inferred from the data fits. This reflects the fact that $\tilde F_0$ does not appear in the expressions for the cosmographic parameters~\eqref{eq_cosmography_fT_FINAL}.
Notice that we have introduced the parameter $\tilde\Omega_m$ in (\ref{eq_definitions_fT_FINAL}), accounting for the correct definition of the matter (baryons plus cold dark matter) density parameter, despite depending on $\tilde F_1$. The reason is obviously that $\tilde F_1$ acts as a rescaling factor for the Newton constant $G_{N,\rm eff}$ = $G_{N,\rm bare}/F_1$, so what is actually probed by cosmological observations is $G_{N,\rm bare}\,\Omega_m = G_{N,\rm eff}\,\tilde\Omega_m$. Solar system tests put strict bounds on deviations of $G_{N,\rm eff}$ from the value of $G_N$ measured in Earth-bound experiments, but as is widely known, for $\FT$ and $f(R)$ theories (see section~\ref{f(R) theories}) $G_{N,\rm eff}$ turns out to be a function of density, so we cannot trivially extend Solar system bounds to cosmological scales and densities. As we have just shown, the cosmographic history of the Universe is unaffected by changes in $F_1$, provided that all other parameters are rescaled accordingly.

Interestingly, rescaling all parameters -- including $\Omega_m$ --  by an additional factor $(1+2\tilde F_2)$, such a factor 
disappears from the denominators in~\eqref{eq_cosmography_fT_FINAL} which leads to particularly simple expressions for $q_0$, $j_0$ and $s_0$. However, we would lose the direct physical interpretation of $\tilde\Omega_m$, so we use the definitions~\eqref{eq_cosmography_fT_FINAL} even though they are slightly more complicated.

\subsection{\texorpdfstring{$f(R)$}{f(R)} Theories} \label{f(R) theories}

As a third example, we consider theories of gravity which are derived from the gravitational action \cite{ReviewsfR3, ReviewsfR4, Capozziello:2011et}:
\begin{eqnarray}
\label{lagr f(R)}
\mathcal{S}\,=\,\int \text{d}^4 x \sqrt{-g}\left[\frac{1}{16\pi G} f(R)+{\cal L}_{m}\right]\,,
\end{eqnarray}
%
The FLRW field equations, when assuming the so-called metric formalism, reduce to
\begin{subequations}\label{eq_fR_field_eqs}
 \begin{align}
  3f_R H^2  & = 8\pi G\rho_m + \frac{R f_R - f}{2} + 3(1+z)H^2f_{RR}R'\;,\\
  2f_R H H' &  = 8\pi G\rho_m+(1+z)H^2f_{RR}R'  
  + (1+z)H\partial_z[(1+z)H f_{RR}R']\;.
 \end{align}
\end{subequations}
where we have used $z$ as independent variable and defined $f_R \equiv \partial f/\partial R$ and analogously for higher derivatives.
Thus, introducing the following definitions,
\begin{subequations}
\label{eq_definitions_fR}
\begin{align}
 \alpha         &= \left. f_R\right|_{z=0}\;,\\
 \beta^2        &= \frac{6H_0^2}{\alpha}\left. f_{RR}\right|_{z=0}\;,\\
 \tilde\Omega_m &= \frac{\Omega_m}{\alpha}\;,\\
 \tilde f_0     &= \frac{1}{6H_0^2\alpha}\left. f\right|_{z=0} - 1 + \frac{\Omega_m}{2}\;,\\
 \tilde f_1     &= \frac{1}{6H_0^2\alpha}\left.\frac{\partial f}{\partial z}\right|_{z=0} - \frac{3\Omega_m}{2}\;,\\
 \tilde f_2     &= \frac{1}{6H_0^2\alpha}\left.\frac{\partial^2 f}{\partial z^2}\right|_{z=0} - 3\Omega_m\,,
\end{align}
\end{subequations}
and using~(\ref{eq_fR_field_eqs}) we find:
\begin{subequations}\label{eq_cosmography_fR}
 \begin{align}
  q_0 &= -1 + \frac{3\tilde\Omega_m}{2} - \tilde f_0 + \beta^2\left(\tilde f_1 + \frac{3\tilde\Omega_m}{2}\right)\,,\\
  j_0 &= 1 - \tilde f_0 - \tilde f_1 + \beta^2\left(\tilde f_1 + \frac{3\tilde\Omega_m}{2}\right)\,,\\
  s_0 &= 1 - \frac{9\tilde\Omega_m}{2} - \tilde f_0^2 + \tilde f_1 + \tilde f_2 + \frac{3\tilde f_1\tilde \Omega_m}{2}  
  + \tilde f_0\left(6 - \tilde f_1 + \frac{3\tilde \Omega_m}{2}\right) - \frac{\beta^4(2\tilde f_1 + 3\tilde\Omega_m)^2}{4}\,  \notag \\ &\quad
  - \frac{\beta^2}{2}(6-2\tilde f_0 + 3\tilde\Omega_m)(2\tilde f_1 + 3\tilde\Omega_m)\,.
 \end{align}
\end{subequations}
Notice that like the case of $F_1$ for $\FT$ theories, the first derivative $\alpha\equiv f_{R,0}$ has disappeared from the expressions. In fact, it simply corresponds to a rescaling of the Newton's constant.
As in the previous cases, we have defined~(\ref{eq_definitions_fR}) in such a way that $\Lambda$CDM corresponds to $\tilde f_i$'s and $\beta$ equal to zero.

Notice that for $f(R)$ theories the mapping between cosmographic and model parameters is not bijective:
\be
(q_0,j_0,s_0) \quad\leftrightarrow\quad (\tilde f_0, \tilde f_1, \tilde f_2, \beta)\,.
\ee
This is a general feature of theories with extra degrees of freedom, as is the case for $f(R)$ gravity (one extra scalar).

\section{Datasets}
We perform cosmological fits to the various theories described above using several low-redshift datasets: SNIa luminosity distance, $H(z)$ and BAO measurements.
These measurements are independent and uncorrelated, thus the total likelihood is taken to be the product of the individual likelihoods, i.e.,
\be
\mathcal L_{\rm tot} = \mathcal L_{\rm SNIa} \times \mathcal L_{H(z)} \times \mathcal L_{\rm BAO}\,,
\ee
and each likelihood is defined as proportional to the exponential of the corresponding $\chi^2$:
\be
\mathcal L_i = \exp\left(-\chi^2_i/2\right)\qquad i=\left\{{\rm BAO},H(z),{\rm SNIa}\right\}\,.
\ee
\subsection{Type Ia Supernovae}
We use SNIa luminosity distance measurements collected in the Union2.1 catalogue~\cite{Suzuki:2011hu}. It contains 580 sources at redshifts $z \le 1.414$, analysed with the SALT-II lightcurve fitter. The distance modulus $\mu \equiv m-M$ is the difference between the observed and absolute magnitude of the object. For a homogeneous, isotropic and spatially flat Universe, it is given by
\be
\mu(z;\bm\theta) = m(z;\bm\theta) - M = 5\log\frac{d_L(z;\bm\theta)}{10\,\rm pc}\,,
\ee
with the luminosity distance
\be\label{eq_SN_luminosity_distance}
d_L(z;\bm\theta) = \frac{(1+z)}{H_0}\int^z_0 \frac{d\zeta}{H(\zeta;\bm\theta)/H_0}\,.
\ee
We denote with $\bm\theta$ all cosmological and model parameters other than $z$. SNIa are standard (or at least standardisable) candles and thus their absolute magnitude is assumed to be constant for each supernova, although its (unknown) value is completely degenerate with $H_0$. Indeed, when fitting cosmological data one should marginalise over the \textit{nuisance parameter} $\Delta_M$ which depends on $H_0$ and $M$, defined by
\be
\mu^{\rm fit}(z_i) = \mu^{\rm Union2.1}(z_i) + \Delta_M\,.
\ee
In the fits, we assume a wide flat prior\footnote{Best fits are of the order of $\Delta M \sim 0.05$, with similar 2--$\sigma$ errors.}
\be
\Delta_M = {\rm Uniform}(-2,2)\,.
\ee
Because we are dealing with fully generic models, $H(z;\bm\theta)$ is not known a priori and we cannot use~\eqref{eq_SN_luminosity_distance} but rather~\eqref{distanzadilumin1e2}. This makes our analysis less accurate because of the finite expansion order, but also more generic because we do not assume any specific model within a given class of theories.

The total $\chi^2$ for SN data is
\be
\chi^2_{\rm SNIa} = \sum_{i=1}^{580} \left[\frac{\mu_i^{\rm fit} - \mu^{\rm th}(z_i;\bm{\theta})}{\sigma_i^2}\right]^2\,,
\ee
where $\mu^{\rm th}(z;\bm\theta)$ holds for the theoretical value of the distance modulus at a given redshift and for a given combination of model and cosmological parameters $\bm\theta$.
\subsection{The Hubble Rate compilation}
The second dataset we consider is the list of Hubble rate measurements at different redshifts. We employ the most recent compilation~\cite{Farooq:2012ju, Farooq:2013hq}, which takes into account 28 pairs of $(H_i,\,z_i)$ with associated errors.

This catalogue represents a novel approach to tracking the Universe's expansion history, providing massive early type galaxies as cosmic chronometers. In fact, the procedure for evaluating $(H_i,\,z_i)$ relies on estimating the differential time $\frac{dt}{dz}$ through different astronomical measurements on galaxies, and then comparing these measurements with the cosmological redshifts at which such galaxies are located. The key relation is
\begin{equation}\label{defdiff}
\frac{{\rm d}z}{{\rm d}t}=-(1+z)\,H(z)\,,
\end{equation}
which allows us to infer $H(z)$ in the right-hand side, once $\frac{dz}{dt}$ and $z$ are independently measured. For our analyses, we use the 28 $H(z)$ measurements reported in ref.~\cite{Farooq:2013hq}, with the list of data spanning in the redshift interval $z\in[0.09,2.30]$.

\subsection{Baryon Acoustic Oscillations}
\label{sec_BAO}

\begin{table}[t]
\begin{minipage}{.55\textwidth}
\centering
\small
\begin{tabular}{c c c c c}
 Survey & $z$ & $d_z$ & Ref.
\\
\hline
\vspace*{-\mytablen}
\\
6dFGS &  0.106 &  $0.3360\pm0.0150$ & \cite{Beutler:2011hx}\\
MGS  &  0.15 & $0.2239\pm 0.0084$ & \cite{Ross:2014qpa}\\
BOSS LOWZ & $0.32$ & $0.1181\pm0.0024$  &\cite{Anderson:2013zyy}\\
SDSS(R) & $0.35$ & $0.1126\pm 0.0022$  &\cite{Padmanabhan:2012hf}\\
BOSS CMASS & $0.57$ & $0.0726\pm0.0007$  &\cite{Anderson:2013zyy}\\
\vspace*{-\mytablen}\\
\hline
\end{tabular}
\caption{BAO $d_z$ data. The measurements come from several experiments as indicated, and are uncorrelated.}
\label{table_BAO_dz}
\end{minipage}
\hfill
\begin{minipage}{.4\textwidth}
\centering
\small
\begin{tabular}{c c}
$z$ & $A(z)$
\\
\hline
\vspace*{-\mytablen}
\\
0.44  &$0.474 \pm 0.034$ \\
0.60  & $0.442\pm 0.020$ \\
0.73  & $0.424 \pm 0.021$ \\
\vspace*{-\mytablen}\\
\hline
\end{tabular}
\caption{BAO WiggleZ data~\cite{Blake:2012pj}. The relative covariance matrix is reported in~\eqref{eq_WiggleZ_covariance}.}
\label{table_BAO_WiggleZ}
\end{minipage}
\end{table}

Baryon Acoustic Oscillations (BAO) represent a typical correlation scale in the matter distribution, and they indeed represent an excellent tool to probe the cosmological history, particularly a combination of the angular diameter distance and the redshift separation. In fact, the typical BAO observable is
\be
d_z(\bm\theta) \equiv \frac{r_{\rm s}(z_{\rm drag})}{D_V(z;\bm\theta)}\,,
\ee
where $r_{\rm s}(z_{\rm drag})$ is the comoving sound horizon at the drag epoch, and
\be
\label{eq_DV_definition}
D_V^3(z) \equiv \frac{z\,d^2_L(z;\bm\theta)}{(1+z)^2H(z;\bm\theta)}\,.
\ee
is the volume-averaged distance, see e.g.~\cite{Eisenstein:2005su}. The quantity $r_{\rm s}(z_{\rm drag})$ must be calibrated assuming a fiducial cosmological model, the Planck data giving~\cite{Ade:2015xua}
\be\label{eq_rs_Planck}
\begin{aligned}
 z_{\rm drag} = 1059.62\pm 0.31\,,\;\;\;
 r_{\rm s}(z_{\rm drag}) = 147.41\pm 0.30\,.
\end{aligned}
\ee
However, such a calibration is impossible without assuming a particular cosmology until redshifts $z\gtrsim 10^3$. This clearly clashes with the idea of keeping our analysis fully general. Although not fully consistent, the Planck value should be safely applicable to any model which does not depart too drastically from $\Lambda$CDM until much lower redshifts, and in which (linear) structure growth is not too severely modified (see also the discussion in section~\ref{S2:Subsection2: Cosmography beyond LCDM}). For this reason, we will assume the Gaussian prior~\eqref{eq_rs_Planck} throughout our analysis.

In addition to the $d_z$ data, we also use data from the WiggleZ collaboration~\cite{Blake:2012pj}. The observable best suited to cosmological fits is in this case $A(z)$, defined by
\be
A(z) = \frac{100 D_V(z)\sqrt{\tilde\Omega_m h^2}}{c\,z}\,,
\ee
where $h \equiv H_0/(100\,{\rm km}\,{\rm s}^{-1}{\rm Mpc}^{-1})$. This quantity is independent of $H_0$, because $D_V\sim H_0^{-1}$.

At this stage let us stress that in $\FT$ and $f(R)$ theories, as mentioned above, the relevant parameter is actually $\tilde\Omega_m$ and not the ``bare'' $\Omega_m$. In the case of quintessence theories, $\tilde\Omega_m = \Omega_m$.

The WiggleZ data are shown in Table~\ref{table_BAO_WiggleZ}. These data points are correlated, with covariance matrix:\footnote{Only the upper half of the symmetric covariance matrix is shown.}
\be
\label{eq_WiggleZ_covariance}
C^{-1} =
\begin{pmatrix}
 1040.3 & -807.5 & 336.8\\
  & 3720.3 & -1551.9\\
  & & 2914.9\\
\end{pmatrix}\,.
\ee
Accordingly the expression for the $\chi^2$ must be suitably modified to include the correlation in (\ref{eq_WiggleZ_covariance}). Thus the total $\chi^2$ for BAO data becomes
\be
\chi^2_\text{BAO} = \chi^2_{d_z} + \chi^2_\text{WiggleZ}\,,
\ee
with
\be
\begin{aligned}
 \chi^2_{d_z} = \sum_{i=1}^5 \left[\frac{{d_z}_i^\text{obs} - d_z^\text{th}(z_i)}{\sigma_d}\right]^2\,,\;\; 
 \chi^2_\text{WiggleZ} = (\mathbf A^\text{obs} - \mathbf A^\text{th})^T C^{-1} (\mathbf A^\text{obs} - \mathbf A^\text{th})\,.
\end{aligned}
\ee

\section{Numerical results}
\label{S4a:results}

\subsection{Fitting Procedure}
\begin{table}[t]
\centering
\small
\def\arraystretch{1.4}
\setlength{\tabcolsep}{.5em}
 \begin{tabular}{r l}
  Parameter   & Prior \hfill\hfill\\
  \hline
  $\Omega_m$                & Uniform(0,1)\\
  $h_0$                     & Uniform(20,120)\\
  $q_0$                     & Uniform(-5,5)\\
  $j_0$                     & Uniform(-20,20)\\
  $s_0$                     & Uniform(-100,100)\\
  $\Delta_M$                & Uniform(-2,2)\\
  $r_{\rm s}(z_{\rm drag})$ & Normal(147.41,0.30)\\
  \hline
\end{tabular}
\caption{Priors for cosmological parameters. The prior on $r_{\rm s}(z_{\rm drag})$ is the Planck value~\eqref{eq_rs_Planck}. The flat priors on ($q_0,j_0,s_0$) result in flat priors on model parameters for quintessence and $f(R)$, but a non-uniform prior on $\tilde F_2$ (see text for details).
}
\label{tab_priors}
\end{table}

We fit our models using each dataset individually and then combine all three datasets, assuming flat priors on cosmological parameters, with the exception of $r_{\rm s}(z_{\rm drag})$ as given in eq.~\eqref{eq_rs_Planck}, for which we have used a Gaussian prior at the Planck best value (see Table~\ref{tab_priors}, and section~\ref{sec_BAO} for further details).
Note that $\Delta_M$ only enters the SNIa analysis where it is completely degenerate with $h_0$; we therefore are unable to constrain it using any dataset individually. For the same reason, $h_0$ is unconstrained by SNIa data unlike in the case of BAO and $H(z)$ measurements. When combining all data, we are able to put constraints on both $h_0$ and $\Delta_M$.

For both quintessence and $f(R)$ theories, the choice of flat priors on $(q_0,j_0,s_0)$ is equivalent to flat priors on the theory parameters $(\tilde V_0,\tilde V_1,\tilde V_2)$ and $(\tilde f_0, \tilde f_1, \tilde f_2, \beta)$ respectively. The Jacobians of the transformations~\eqref{eq_cosmography_ST} and~\eqref{eq_cosmography_fR} are simply constants. Since $f(R)$ theories possess one parameter more than the other theories under consideration, we have actually been forced to specify the prior on $\beta$ because it is not implied by table~\ref{tab_priors}; thus we have chosen the wide flat prior
\be
\beta = {\rm Uniform(-100,100)}\,.
\ee
One can easily check a posteriori that the posterior distributions lie well within the assumed priors.

In the case of $\FT$ theories, the dependence between cosmographic and model parameters is less trivial as seen in eqs.~(\ref{eq_cosmography_fT_FINAL}). Then, denoting $Q_i \equiv (q_0,j_0,s_0)$ and $F_i \equiv (\tilde F_0, \tilde F_1, \tilde F_2)$,  we have
\be
\det J_{ij} \equiv \det \frac{\partial Q_i}{\partial F_j} \propto \frac{\tilde\Omega_m^6}{(1+2\tilde F_2)^9}\,.
\ee
Therefore, a flat prior on cosmographic parameters results in a prior on $\tilde F_2$, which has a penalty for large $\tilde F_2$ proportional to $(1+2\tilde F_2)^{-9}$. Inspection of~\eqref{eq_cosmography_fT_FINAL} reveals that at large values of $\tilde F_2$, variations in $\tilde F_2$ have little effect on cosmographic parameters, in particular $q_0 \simeq -1$. When the MCMC chains explore such tails, a flat prior on $\tilde F_2$ may result in an almost unconstrained progression towards $\tilde F_2\to \infty$ with little increase in the volume of the effective parameter space, i.e., the cosmographic parameter space, being explored. This problem is avoided choosing flat priors on cosmographic parameters.

We should emphasise that while we choose flat priors on $(q_0,j_0,s_0)$, our fits use the model parameters directly, following the prescriptions~(\ref{eq_cosmography_ST}, \ref{eq_cosmography_fT_FINAL}, \ref{eq_cosmography_fR}), with a completely independent analysis for each model.

Finally, let us mention that we have firstly considered only the data at redshifts $z<1$, which are 22 and 551 data points for $H(z)$ and SNIa measurements, respectively and have then repeated the analysis for the full datasets. This is to check the consistency of using a finite expansion in $z$ at somewhat large redshifts $z<1$. For BAO, the full dataset are contained in  $z<1$, so the same data will be used in both fits.
We have run MCMC fitting codes equipped with a Gelman-Rubin convergence diagnostics, and performed the statistical analysis using publicly available \texttt{Python} codes.\footnote{\url{http://getdist.readthedocs.org/en/latest/}.}

\subsection{Exact vs. Truncated \texorpdfstring{$\Lambda$}{Lambda}CDM}

\begin{figure}[t]
\begin{minipage}{.58\textwidth}
\centering
\small
\setlength{\tabcolsep}{1.5pt}
\def\arraystretch{1.4}
\begin{tabular}{c c c}
 \textbf{Parameter}  &$\bm{z<1}$                      & \bfseries All--$\bm z$\\
 \hline
 & \multicolumn{2}{c}{\bfseries Exact $\bm\Lambda$CDM}\\
 \hline
 $h_0$               & $69.1^{+1.4}_{-1.4}(69.1)$           & $69.3^{+1.2}_{-1.2}(69.3)$         \\
 $\Delta_M$          & $-0.023^{+0.037}_{-0.041}(-0.028)$   & $-0.019^{+0.035}_{-0.035}(-0.015)$ \\
 $\Omega_m$          & $0.291^{+0.037}_{-0.035}(0.289)$     & $0.283^{+0.025}_{-0.023}(0.283) $  \\
 $\chi^2_{\rm min}$  & 552.8/(581 d.o.f.)                   & 584.5/(616 d.o.f.)       \\
 \hline
 & \multicolumn{2}{c}{\bfseries Truncated $\bm\Lambda$CDM}\\
 \hline
$h_0$              & $69.5^{+1.4}_{-1.4}(69.3)$         & $70.2^{+1.5}_{-1.5}(70.1)$             \\
$\Delta_M$         & $-0.014^{+0.040}_{-0.040}(-0.021)$ & $-0.003^{+0.040}_{-0.040}(-0.0054)$    \\
$\Omega_m$         & $0.278^{+0.034}_{-0.032}(0.278)$   & $0.248^{+0.031}_{-0.028}(0.240)$       \\
$\chi^2_{\rm min}$ & 552.2/(581 d.o.f.)                 & 601.5/(616 d.o.f.) \\
 \hline
 \end{tabular}
 \captionof{table}{Summary of results for the exact and truncated $\Lambda$CDM models. Reported correspond to 95\% confidence levels; best-fit values are in brackets.}
 \label{tab_results_LCDM}
\end{minipage}
\hfill
\begin{minipage}{.4\textwidth}
\centering
\vspace{0pt}
\includegraphics[width=.95\textwidth]{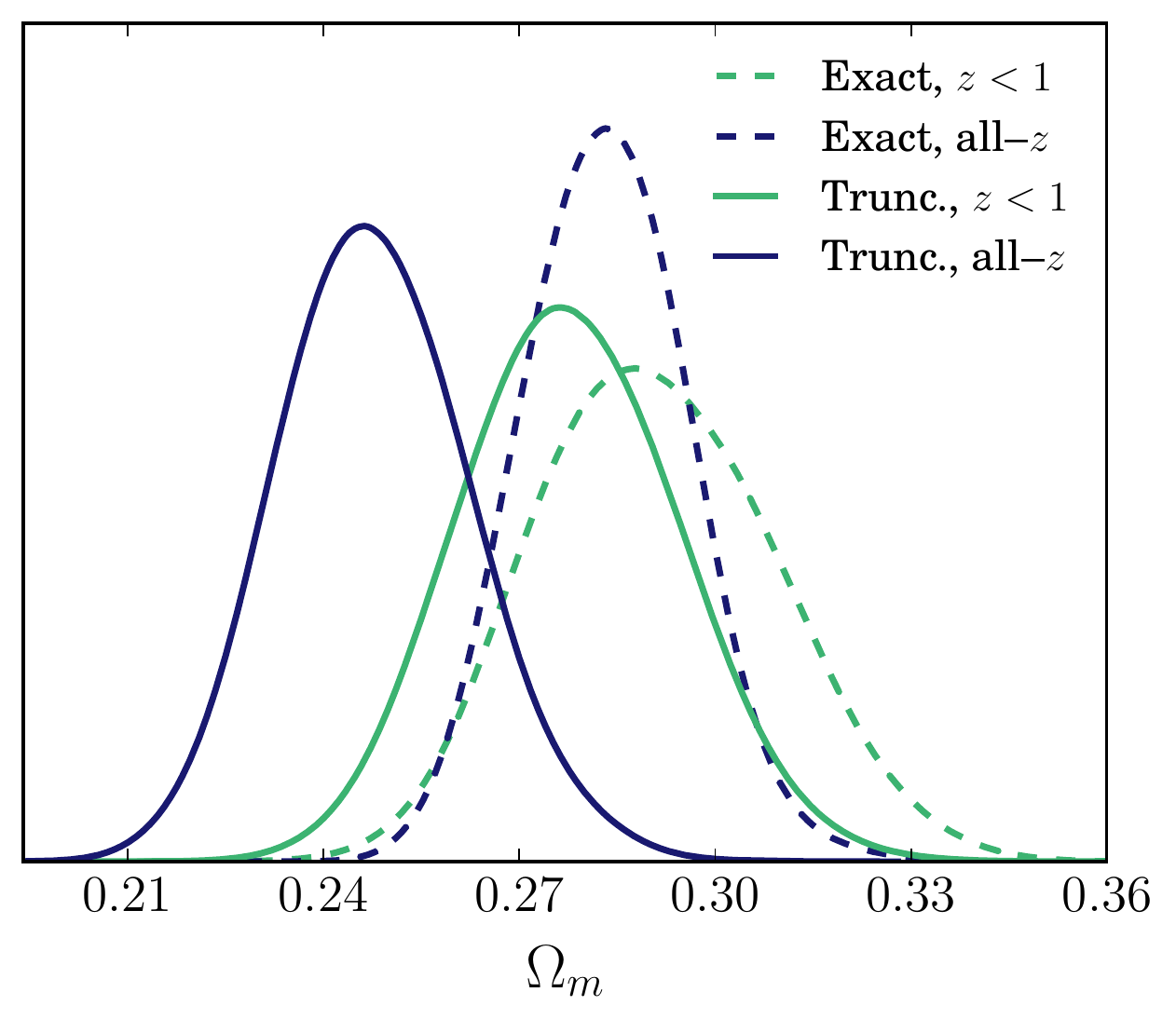}
\caption{Comparison between exact and truncated $\Lambda$CDM models. The inferred values for $\Omega_m$ are essentially compatible at roughly 1-$\sigma$. We notice that the truncated model gives slightly lower values of $\Omega_m$ than the exact one, and that $z<1$ give slightly larger values of $\Omega_m$ compared to the all--$z$ data.}
\label{fig_results_LCDM}
\end{minipage}
 \end{figure}

For a straightforward comparison with modified gravity theories, let us first present the results for $\Lambda$CDM, for which the only free parameter is $\Omega_m$.
Note that the approach one takes when fitting data to $\Lambda$CDM is intrinsically different than other cosmological theories, because for $\Lambda$CDM an exact solution $H(z)$ is at hand and therefore also analytical expressions for all cosmological observables, which can be integrated numerically for any $z$ and parameter combination.
For the other theories studied in this paper, as explained in the previous sections, a finite expansion around $z=0$ must be considered in order to keep our analysis as general as possible. Results in Table \ref{tab_results_LCDM} show
the existence of an excellent agreement between $z<1$ and all--$z$ data, with slightly smaller $\Omega_m$ in the all--$z$ fits.

It might be instructive to see how results change if we treat $\Lambda$CDM in the same way as the other theories, namely expanding using the relations:
\begin{subequations}
\label{eq_LCDM_q_j_s}
\begin{align}
 q_0 = -1 + \frac{3\Omega_m}{2}\,,\;\;
   j_0 = 1\,,\;\;
   s_0 = 1 - \frac{9\Omega_m}{2}\,,
\end{align}
\end{subequations}
which describe the expansion of $\Lambda$CDM around $z=0$.
By doing so, we can test the consistency of our method and also investigate if and how constraints are affected by truncating the expansion of $H(z)$ at a finite order. In other words, comparing results for the exact $\Lambda$CDM model and for this truncated $\Lambda$CDM can help us estimate the importance of high-order corrections to the cosmographic expansion, and for which parameters we expect them to be most relevant.

Our results for the truncated $\Lambda$CDM model are shown in Table~\ref{tab_results_LCDM}, where one can see how %
the exact and truncated models agree within 1--$\sigma$ confidence level, the agreement being better for $z<1$ data than for the all--$z$ dataset. This is somewhat expected, since high-order terms should be more relevant at higher redshifts, where the two models are more likely to differ.
Interestingly, $\Omega_m$ tends to be larger for the the exact $\Lambda$CDM model. In terms of $q_0,j_0,s_0$~\eqref{eq_LCDM_q_j_s}, this translates into larger values of $q_0$ and smaller values of $s_0$. Errors are essentially the same in both cases, which indicates a negligible gain in precision when using the exact model instead of the truncated expansion. However, there seems to be a bias effect due to the finiteness of the expansion, which leads to lower values of $\Omega_m$ and in turn to a smaller $q_0$ and a larger $s_0$.

\subsection{Quintessence Theories}\label{sec_results_kessence}

\begin{figure}[t]

\begin{minipage}{.44\textwidth}
\centering
\vspace{0pt}
\includegraphics[width=1\textwidth]{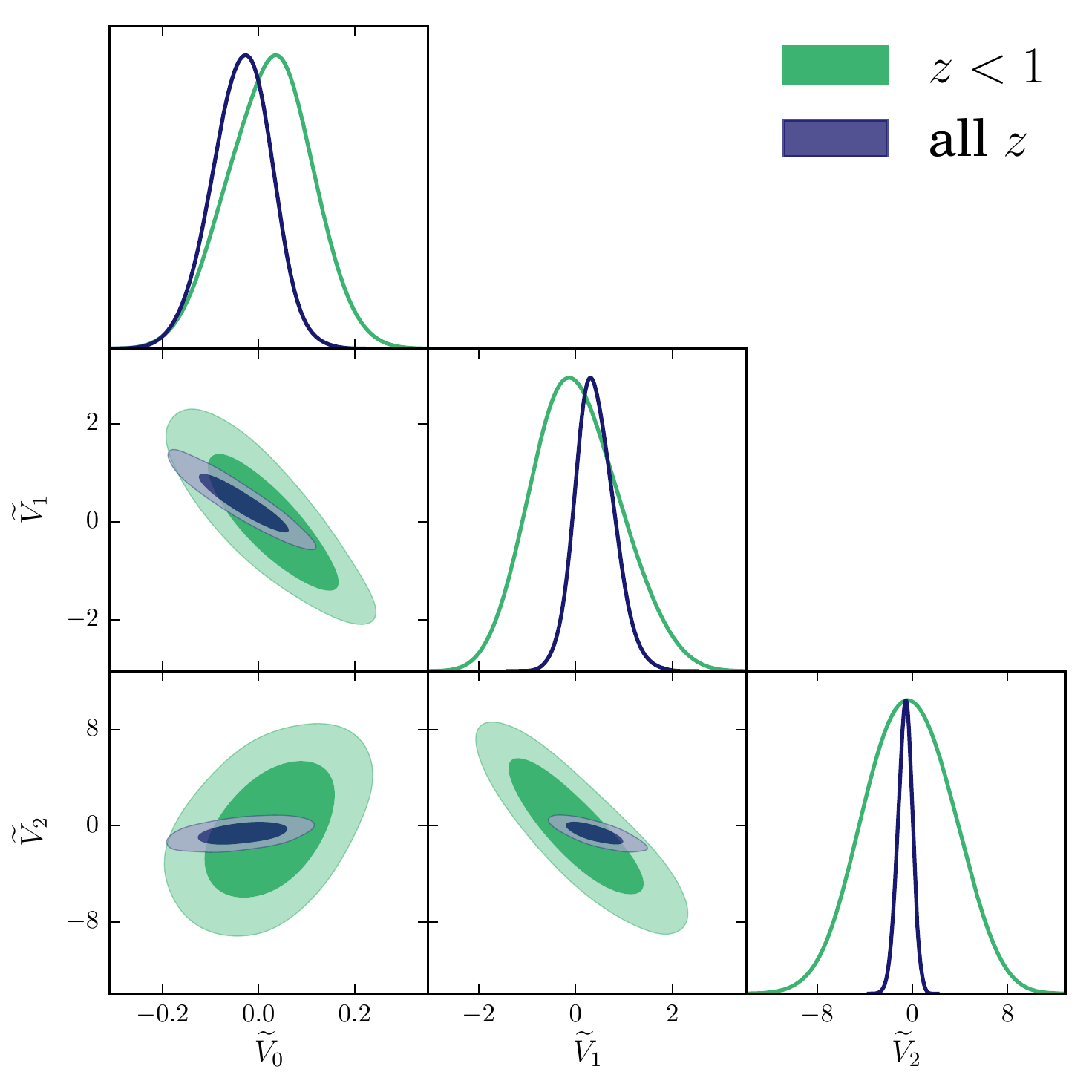}
\end{minipage}
\hfill
\begin{minipage}{.55\textwidth}
\centering
\small
\setlength{\tabcolsep}{1.5pt}
\def\arraystretch{1.4}
\begin{tabular}{c c c}
 \textbf{Parameter}  &$\bm{z<1}$                      & \bfseries All--$\bm z$\\
 \hline
 $h_0$              & $69.8^{+2.1}_{-2.0}(70.1)$           & $69.4^{+1.8}_{-1.8}(69.5) $\\
 $\Delta_M$         & $-0.007^{+0.050}_{-0.047}(-0.002)$   & $-0.011^{+0.043}_{-0.045}(-0.012)$ \\
 $\Omega_m$         &$0.296^{+0.066}_{-0.061}(0.273)$      & $0.294^{+0.049}_{-0.045}(0.295)$\\
 $\tilde V_0$       &$0.02^{+0.17}_{-0.17}(0.0477) $       & $-0.04^{+0.12}_{-0.13}(0.0061)  $\\
 $\tilde V_1$       &$0.0^{+1.9}_{-1.7}(-0.467)      $     & $0.41^{+0.86}_{-0.76}(0.126)   $\\
 $\tilde V_2$       &$-0.3^{+7.0}_{-6.9}(1.60)     $       & $-0.6^{+1.2}_{-1.3}(-0.364)     $\\
 $\chi^2_{\rm min}$ & 551.8/(581 d.o.f.)                   & 585.0/(616 d.o.f.)       \\
\hline
\end{tabular}
 \captionof{table}{
 Summary of results for quintessence theories. Parameters are defined in~\eqref{eq_definitions_ST}, and reported errors are 95\% confidence level; best fit values are shown in brackets.}
 \label{tab_results_k_essence}
\end{minipage}
\caption{Results for quintessence theories, see Table~\ref{tab_results_k_essence}. All model parameters~\eqref{eq_definitions_ST} are compatible with zero at 1-$\sigma$, indicating that $\Lambda$CDM is the favoured limit in this case. Nonetheless, a relevant portion of parameter space is still allowed by the data, see also section~\ref{sec_results_kessence}}.
\label{fig_kessence_results}
\end{figure}

Our results for quintessence theories are summarised in table~\ref{tab_results_k_essence} and figure~\ref{fig_kessence_results}.
All model parameters $\tilde V_i$ are compatible with zero within 2--$\sigma$ confidence level in all cases, which supports the evidence for $\Lambda$CDM as an excellent approximation for the cosmological expansion history. On the other hand, the 1--$\sigma$ and 2--$\sigma$ confidence level regions allow for some deviation, with $\tilde V_0$ constrained at $\sim \mathcal O(0.1)$, and $\tilde V_1$ and $\tilde V_2$ constrained at the level of~$\mathcal O(1)$.
As one can see in figure~\ref{fig_kessence_results}, the fits with $z<1$ and for all $z$ agree perfectly, and the contribution of $z>1$ data is, as expected, more evident for parameters involving higher derivatives. The mean values agree within 1--$\sigma$ confidence level, and in the full analysis the errors are reduced by a factor between 1.5 (for $\Omega_m$) and roughly 5 (for $\tilde V_2$).
For definiteness, let us consider a simple model with a quadratic potential as follows
\be
V(\phi(z))\, = \,\frac{3H_0^2}{8\pi G}\Omega_z + \frac{1}{28\pi G}m_z^2(z-\zeta)^2\,,
\ee
where $\Omega_z$ and $\zeta$ are dimensionless constants, and $m_z$ can be understood as the mass associated to the scalar field.\footnote{The factor $8\pi G$ in the denominator of the mass term is present for dimensional reasons: scalar fields have conventionally natural dimensions of [Energy] whereas $z$ is dimensionless.} For this model we have
\be
 \tilde V_0 = \frac{m_z^2\zeta^2}{2H_0^2} - 3(1 - \Omega_m - \Omega_z)\,\,\,,\,\,\,
 \tilde V_1 = -\frac{m_z^2\zeta}{H_0^2}\,\,\,,\,\,\,
 \tilde V_2 = \frac{m_z^2}{H_0^2}\,.
\ee
The constraints $|\tilde V_0| \lesssim 0.1$, $|\tilde V_{1,2}|\lesssim 1$ provide:
\be
\begin{aligned}
 \left|m_z\right|\lesssim H_0\,,\;\;
 \left|\zeta\right| \lesssim 1\,,\;\;
 \left|1 - \Omega_z - \Omega_m\right| \lesssim 0.2\,.
\end{aligned}
\ee
Let us comment on these values. The mass $m_z$ of the scalar field must be at most of the order of the present Hubble rate, which is precisely what would be expected for scalar fields responsible for the present cosmic acceleration. The constant term $\Omega_z$ is very close to $1-\Omega_m$, which would be the case for a flat $\Lambda$CDM model identifying $\Omega_z \to \Omega_\Lambda$.

\subsection{\texorpdfstring{$\FT$}{F(T)} Theories}

\begin{figure}[tb]
\begin{minipage}{.44\textwidth}
 \centering
 \vspace{0pt}
\includegraphics[width=1\textwidth]{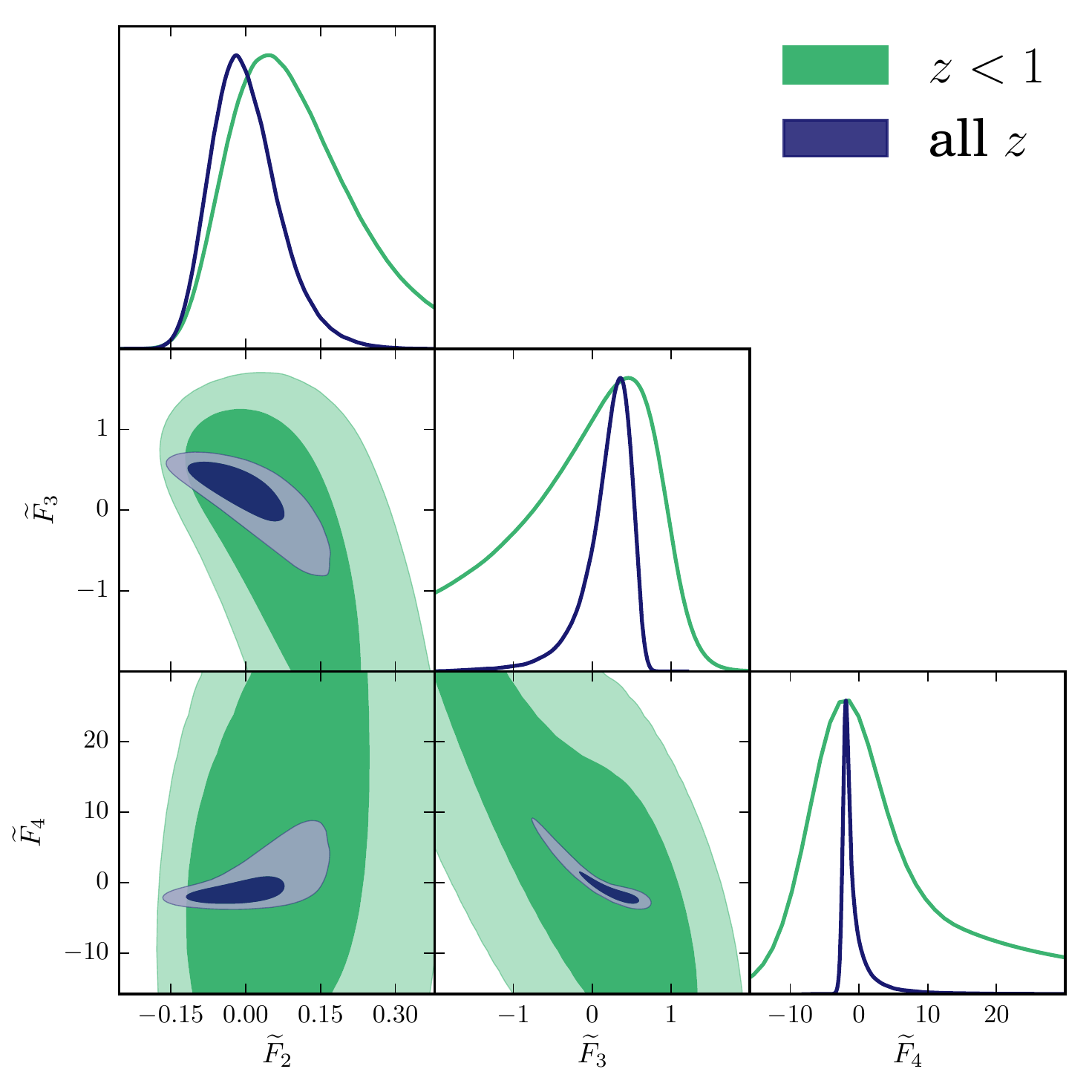}
\end{minipage}
\hfill
\begin{minipage}{.55\textwidth}
 \centering
 \small
\setlength{\tabcolsep}{1.5pt}
\def\arraystretch{1.4}
\begin{tabular}{c c c}
 \textbf{Parameter}    &$\bm{z<1}$                             & \bfseries All--$\bm z$\\
\hline
$h_0$                  & $70.1^{+1.9}_{-1.9}(69.2)$          & $69.8^{+1.8}_{-1.8}(69.2)$         \\
$\Delta_M$             & $-0.002^{+0.047}_{-0.047}(-0.023)$  & $-0.004^{+0.044}_{-0.044}(-0.020)$ \\
$\widetilde\Omega_m$   & $0.294^{+0.062}_{-0.058}(0.294)$    & $0.294^{+0.052}_{-0.047}(0.289)$   \\
$\widetilde F_2$       & $0.11^{+0.28}_{-0.23}(0.029)$       & $0.00^{+0.14}_{-0.13}(-0.024)$     \\
$\widetilde F_3$       & $-1.3^{+2.7}_{-5.4}(0.028)$         & $0.18^{+0.49}_{-0.68}(0.293)$      \\
$\widetilde F_4$       & $42^{+200}_{-70}(-0.66)$            & $-0.5^{+5.6}_{-2.7}(-1.636)$       \\
$\chi^2_{\rm min}$     & 552.2/(581 d.o.f.)                  & 585.6/(616 d.o.f.)                 \\
\hline
\end{tabular}
\captionof{table}{Results for $\FT$ theories. Errors correspond to 95\% confidence levels; best fit values are in brackets.}
\label{tab_results_F_T}
\end{minipage}
 \caption{Results for $\FT$ theories for $z<1$ and all--$z$. All model parameters are nicely compatible with zero ($\Lambda$CDM limit), but $\tilde F_3$ and especially $\tilde F_4$ can deviate rather substantially from zero, particularly for the $z<1$ analysis.}
 \label{fig_results_F_T}
\end{figure}

Our results for $\FT$ theories are shown in Table~\ref{tab_results_F_T} and figure~\ref{fig_results_F_T}.
All model parameters are compatible with zero at about 1--$\sigma$ level in the all--$z$ analysis. Still, the 95\% confidence levels allow for quite a large parameter range, particularly for higher derivatives ($\tilde F_3$, $\tilde F_4$).
This also appears in a rather dramatic way for the $z<1$ results, whereby $\tilde F_3$ and $\tilde F_4$ are only constrained at the level of $|\tilde F_3|\lesssim 5$ and $|\tilde F_4|\lesssim 200$. However, this is presumably merely an indication of the flatness of the $\chi^2$ manifold for varying $\tilde F_i$ for the $z<1$ data, rather than an indication of a true departure from $\Lambda$CDM. The position of the best-fit points, very close to $\tilde F_i = 0$, supports this conclusion.
Interestingly, the posterior probabilities are far from Gaussian, with long tails at large positive or negative values, with
\be
\begin{aligned}
 \tilde F_{2,4} &\gtrsim 0\,\,\,,\,\,\,
 \tilde F_3 &\lesssim 0\,.
\end{aligned}
\ee

\subsection{\texorpdfstring{$f(R)$}{f(R)} Theories}

\begin{figure}[tbh]

\begin{minipage}{.47\textwidth}
 \centering
 \vspace{0pt}
\includegraphics[width=1\textwidth]{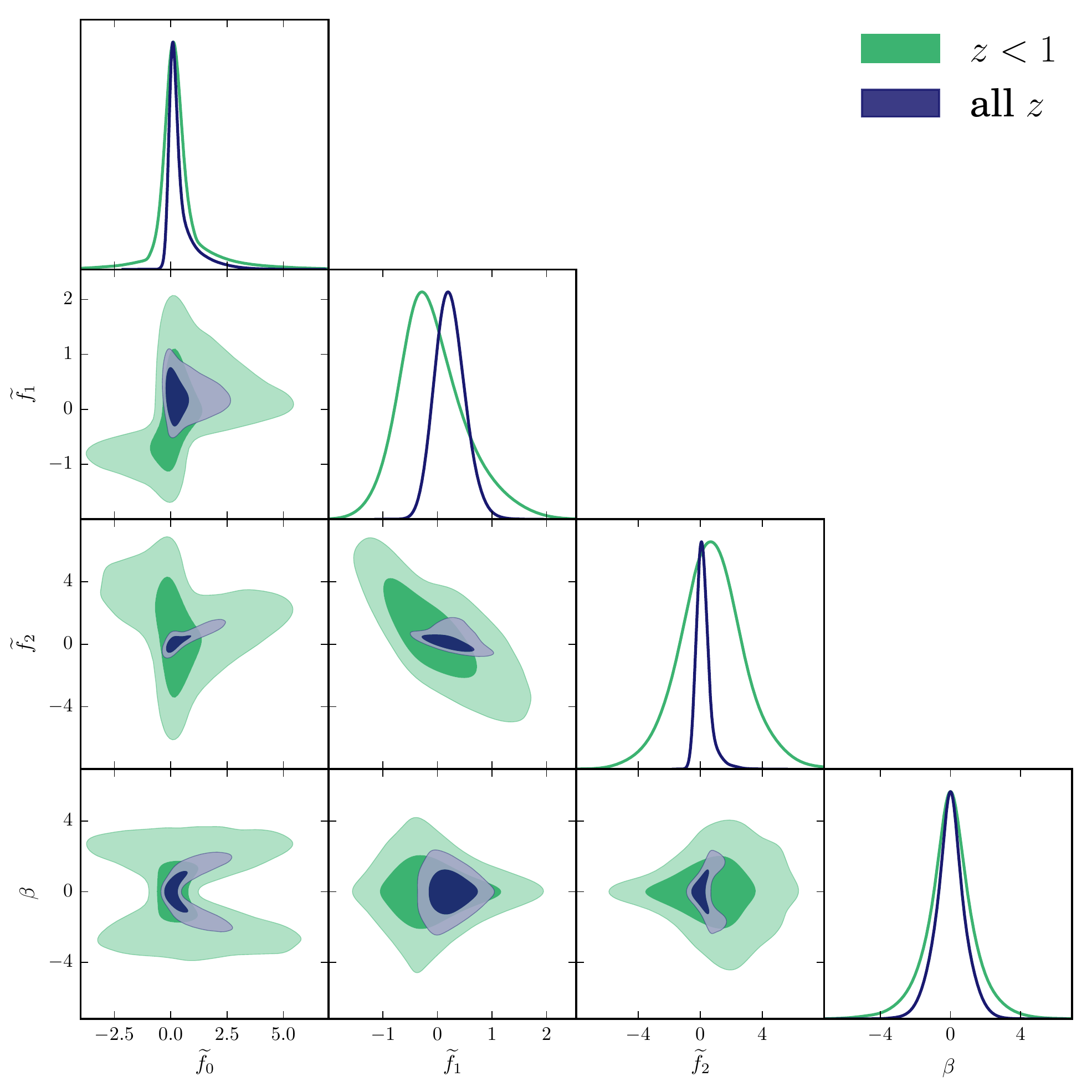}
\end{minipage}
\hfill
\begin{minipage}{.52\textwidth}
 \vspace{0pt}
 \small
\setlength{\tabcolsep}{1.5pt}
\def\arraystretch{1.4}
\begin{tabular}{c c c}
 \textbf{Parameter}  &$\bm{z<1}$                         & \bfseries All--$\bm z$\\
\hline
$h_0$               & $69.9^{+2.0}_{-1.9}(69.6)$         & $69.6^{+1.8}_{-1.8}(69.3)$\\
$\Delta_M$          & $-0.006^{+0.048}_{-0.048}(-0.017)$ & $-0.007^{+0.045}_{-0.045}(-0.018)$\\
$\widetilde\Omega_m$& $0.292^{+0.063}_{-0.060}(0.284)$   & $0.293^{+0.053}_{-0.049}(0.292)   $\\
$\widetilde f_0$    & $0.4^{+3.5}_{-2.3}(4.63)$          & $0.40^{+1.5}_{-0.70}(0.348)      $\\
$\widetilde f_1$    & $-0.1^{+1.3}_{-1.1}(-0.017)$       & $0.23^{+0.55}_{-0.53}(0.300)       $\\
$\widetilde f_2$    & $0.7^{+4.5}_{-4.3}(1.95)$          & $0.20^{+1.0}_{-0.86}(0.074)  $\\
$\beta$             & $0.0^{+3.0}_{-3.0}(-3.951)$        & $0.0^{+1.8}_{-1.8}(0.711)$\\
$\chi^2_{\rm min}$  & 552.1/(581 d.o.f.)                 & 585.6/(616 d.o.f.) \\
\hline
\end{tabular}
\captionof{table}{Results for $f(R)$ theories. Errors correspond to 95\% confidence levels; best fit values are in brackets.}
\label{tab_results_f_R}
\end{minipage}
\caption{Results for $f(R)$ theories for $z<1$ and all--$z$. The model parameters are nicely centred around zero, but departures of order unity are allowed. The odd shapes of the contours indicates that the likelihood surface differs from a multi-variate Gaussian even in the vicinity of the maximum likelihood point.}
\label{fig_results_f_R}
\end{figure}

Our results for these theories are shown in Table~\ref{tab_results_f_R} and figure~\ref{fig_results_f_R}.
We see that all model parameters are compatible with zero at about 1--$\sigma$ level. Notably, the contours have shapes quite different from ellipses, which can be understood as an indication of the complexity of the relations between the cosmographic and model parameters, as one can see inspecting eq.~\eqref{eq_cosmography_fR}. Moreover, such contours change rather drastically in shape and size from the $z<1$ to the all--$z$ case, although the higher-likelihood regions are still nicely overlapping. This is most probably an indication of how well $f(R)$ theories are capable of mimicking $\Lambda$CDM: unlike e.g., quintessence theories, where contours are ellipses and change relatively little from $z<1$ to all--$z$ except for their size, $f(R)$ theories parameters are extremely sensitive to changes in the cosmographic parameters, and in a highly non-trivial way. In other words, a small variation in the uncertainties on $q_0$, $j_0$ and $s_0$ might lead to a large and complex variation in the uncertainties on $f(R)$ parameters.

On the other hand, this also means that an increase in the precision in determining the cosmographic parameters might lead to a possibly significant reduction of the parameter space for $f(R)$ theories. Let us consider for instance the regions at $\tilde f_0 \lesssim 0$ in figure~\ref{fig_results_f_R}. Although allowed by the $z<1$ analysis, such a region disappears completely from the all--$z$ fits, despite the small improvement in the determination of $q_0$ in the two cases (as evidenced by the very similar contours for $\Omega_m$ in the $\Lambda$CDM analysis, or $\tilde V_0$ in quintessence theories).

\subsection{Model Selection Criteria}

\begin{table}[t]
\centering
\small
\setlength{\tabcolsep}{2.6pt}
\def\arraystretch{1.4}
 \begin{tabular}{c c c c c c}
  \textbf{Model}  & $\bm{\Delta d}$  & $\bm{\chi^2_{\rm min}}$ & $\bm\Delta$\textbf{AIC} & $\bm\Delta$\textbf{AICc} & $\bm\Delta$\textbf{BIC} \\
\hline
&  \multicolumn{5}{c}{$\bm{z<1}$}\\
\hline
Exact $\Lambda$CDM       &  0    & 552.8  & --     & --      & --       \\
  Trunc.~$\Lambda$CDM    &  0    & 552.2  & -0.60  & -0.60   & -0.60     \\
  quintessence           &  3    & 551.8  & 5.00   & 5.04    & 18.1     \\
  $\FT$                  &  3    & 552.2  & 5.40   & 5.44    & 18.5     \\
  $f(R)$                 &  3    & 552.1  & 5.30   & 5.34    & 18.4     \\
   \hline
   & \multicolumn{5}{c}{\textbf{All--}$\bm z$}\\
   \hline
  Exact $\Lambda$CDM     &  0    & 584.5  & --    & --    & --     \\
  Trunc.~$\Lambda$CDM    &  0    & 601.5  & 17.0  & 17.0  & 17.0   \\
  quintessence           &  3    & 585.0  & 6.50  & 6.54  & 19.8   \\
  $\FT$                  &  3    & 585.6  & 7.10  & 7.14  & 20.4   \\
  $f(R)$                 &  3    & 585.6  & 7.10  & 7.14  & 20.4   \\
  \hline
 \end{tabular}
 \caption{Model selection criteria. Modified gravity theories are strongly disfavoured compared the exact $\Lambda$CDM model, but are actually favoured to the truncated $\Lambda$CDM model if one follows the AIC(c). Unsurprisingly, modified gravity is still disfavoured despite the lower $\chi^2_{\rm min}$ according to the BIC, due to the presence of extra parameters.}
 \label{tab_selection_criteria}
\end{table}

We compare models using several well-known model selection criteria, namely the Akaike Information Criterion (AIC), the corrected Akaike Information Criterion (AICc), and the Bayes Information Criterion (BIC), defined as (see e.g.~\cite{opac:b1100695}):
\be
\begin{aligned}
 & {\rm AIC} \equiv -2\ln \mathcal L + 2d\,,\\
 & {\rm AICc}  \equiv {\rm AIC} + \frac{2d(d+1)}{N-d-1}\,,\\
 & {\rm BIC}  \equiv -2\ln \mathcal L + d\ln N\,,
\end{aligned}
\ee
where $d$ is the number of parameters of the model, and $N$ is the number of data points.

These quantities contain possibly large scaling constants, due especially to the largeness of the number of data points considered. However, the differences
\be
\Delta X = X_{\rm model} - X_{\Lambda\rm CDM}\,,\quad X = {\rm AIC, AICc, BIC}
\ee
are independent of these scaling, and are a measure of the information loss experienced when fitting using a model instead of the reference model, that is exact $\Lambda$CDM. As general rule of thumb, one usually considers $\Delta \leq 2$ to indicate substantial support (evidence), $4\leq \Delta \leq 7$ much less support, and $\Delta \geq 10$ essentially no support.

The results of our analysis are shown in Table~\ref{tab_selection_criteria}. The truncated $\Lambda$CDM model contains no additional parameters with respect to the exact model, with $h_0$, $\Delta_M$ and $\Omega_m$ being the only parameters upon which all observables depend. In particular, the cosmographic parameters $q_0$, $j_0$ and $s_0$ are fixed once we choose $\Omega_m$.

On the other hand, quintessence, $\FT$ and $f(R)$ theories all have three extra parameters. In fact, the independent combinations of parameters in all these three theories turn out to be $q_0$, $j_0$ and $s_0$, regardless of the considered theory. None of these three quantities can be derived simply from $\Omega_m$ as in the $\Lambda$CDM models, as discussed above.

Concerning the truncation, we should make the following comments:
\begin{itemize}

\item For $z<1$ data, the truncated-$\Lambda$CDM model is actually slightly preferred to the exact model, with almost identical $\chi^2_{\rm min}$, whereas the other models are very weakly supported ($\Delta$AIC $\sim 5$). The $\Delta$BIC takes on very large values, due to the much higher penalty attributed by this criterion to extra parameters.

\item For all--$z$ data, the situation is more complicated. The exact $\Lambda$CDM model is strongly preferred compared to the truncated model, whereas the other models have essentially the same $\chi^2_{\rm min}$ as the exact (best) model, which results in very little support in their favour because of the presence of extra parameters. However, there is very strong support for each of the models considered  when compared to the truncated $\Lambda$CDM model, with $\Delta$AIC $\lesssim -10$, but still no support using the BIC, $\Delta$BIC $\sim 9$.
\end{itemize}

On the one hand, all theories under consideration in this paper should of course be compared to the Concordance model (the exact $\Lambda$CDM model), and when doing so all of them appear to be strongly disfavoured. On the other hand, in making this comparison,  we limit ourselves to a finite expansion order for the three theories,  but not for $\Lambda$CDM. If, instead, we use a finite expansion for the Concordance model as well, i.e., the truncated $\Lambda$CDM model, these  alternative theories considered enjoy much stronger support, at least when using the AIC and AICc. In other words, it seems that if we did not know that $\Lambda$CDM is exactly solvable and only expanded it up to third order around $z=0$, as we do for the other models, we would actually find strong support for these alternative theories of gravity.

We stress that we should not take this result too literally and interpret it as ``Bayesian evidence'' for modified gravity. As discussed in~\ref{sec:problems}, we are using a finite order expansion but using data at $z>1$, hence statistical biases very likely plays a considerable role. The apparent ``rejection'' of the truncated $\Lambda$CDM is likely an artifact of these approximations and will definitely disappear at higher orders, because we found that the exact $\Lambda$CDM model is by far the preferred one.

When using a truncated expansion of a model to given order, we are essentially fixing all higher derivatives of the Hubble rate to zero. For $\Lambda$CDM, where derivatives of $H$ depend only on $\Omega_m$, this implies a very tight relation between $q,j,s,\dots$, which are not allowed to vary \textit{independently}. In turn, this leads to a relatively poor fit of cosmological data. For modified theories, the additional parameters result in greater freedom in the relative dependence of $q,j,s,\dots$ and in an overall better fit.

All in all, we can interpret our results as follows. We know that $\Lambda$CDM works extremely well for the complete cosmological expansion history, and particularly so at late times. Therefore, any alternative theory will have to mimic $\Lambda$CDM rather precisely if it is to be compatible with cosmological data. In turn, we do not expect a significant improvement in terms of $\chi^2_{\rm min}$ when considering alternative theories. It is because these theories contain extra parameters that they will in general be disfavoured, very strongly in fact, in the case of the BIC which heavily penalises additional parameters.

\subsection{Comparison with Previous Work}

\begin{table}[t]
\centering
\small
\setlength{\tabcolsep}{2.6pt}
\def\arraystretch{1.4}
\begin{tabular}{c c c}
 \textbf{Parameter}  &$\bm{z<1}$                      & \bfseries All--$\bm z$\\
 \hline
 & \multicolumn{2}{c}{$\bm{\FT,\,F_2 = 0}$}\\
 \hline
 $h_0$              & $69.6^{+1.5}_{-1.6}(69.4)$           & $69.8^{+1.4}_{-1.4}(69.5)         $ \\
 $\Delta_M$         &$-0.011^{+0.043}_{-0.044}(-0.017)$    & $-0.004^{+0.039}_{-0.039}(-0.013) $ \\
 $\Omega_m$         &$0.296^{+0.069}_{-0.063}(0.282)$      & $0.295^{+0.051}_{-0.045}(0.297)   $ \\
 $\tilde F_3$       &$0.2^{+1.1}_{-1.2}(-0.01)$            & $0.23^{+0.31}_{-0.33}(0.225)      $ \\
 $\tilde F_4$       &$1.2^{+18}_{-9.4}(0.726)$             & $-1.2^{+2.4}_{-1.7}(-1.38)        $ \\
 $\chi^2_{\rm min}$ & 552.4/(581 d.o.f.)                   & 585.7/(616 d.o.f.)                  \\
\hline
 & \multicolumn{2}{c}{$\bm{f(R),\,\beta = 0}$}\\
 \hline
 $h_0$              &$69.7^{+2.1}_{-1.9}(69.7)        $    & $69.6^{+1.8}_{-1.8}(69.4)        $\\
 $\Delta_M$         &$-0.009^{+0.050}_{-0.048}(-0.019)  $    & $-0.008^{+0.045}_{-0.045}(-0.012)  $\\
 $\Omega_m$         &$0.297^{+0.068}_{-0.063}(0.284)   $    & $0.294^{+0.051}_{-0.047}(0.288)   $\\
 $\tilde f_0$       &$0.02^{+0.17}_{-0.17}(0.046)      $    & $-0.02^{+0.12}_{-0.12}(-0.023)     $\\
 $\tilde f_1$       &$0.1^{+1.5}_{-1.4}(-0.256)         $    & $0.26^{+0.59}_{-0.52}(0.234)      $\\
 $\tilde f_2$       &$-0.2^{+5.0}_{-4.5}(0.857)        $    & $-0.11^{+0.60}_{-0.58}(-0.072)     $\\
 $\chi^2_{\rm min}$ & 552.1/(581 d.o.f.)                   & 585.6/(616 d.o.f.)          \\
\hline
\end{tabular}
 \caption{
 Summary of results for $\FT$ and $f(R)$ theories, for $(F_1,F_2) = (1,0)$ and $(\alpha,\beta) = (1,0)$ respectively.}
 \label{tab_results_F2_beta}
\end{table}

\begin{figure}[ht]
\centering
\includegraphics[width=.4\textwidth]{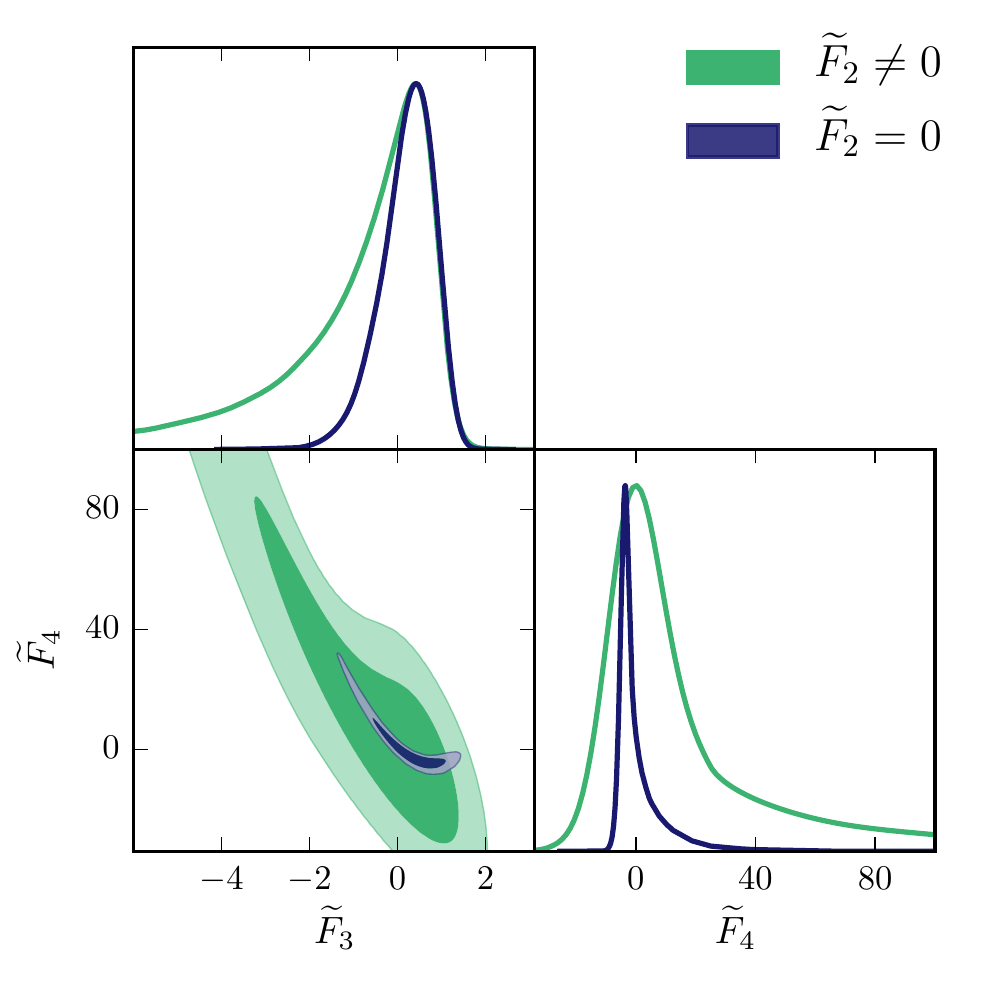}
\hspace{3em}
\includegraphics[width=.4\textwidth]{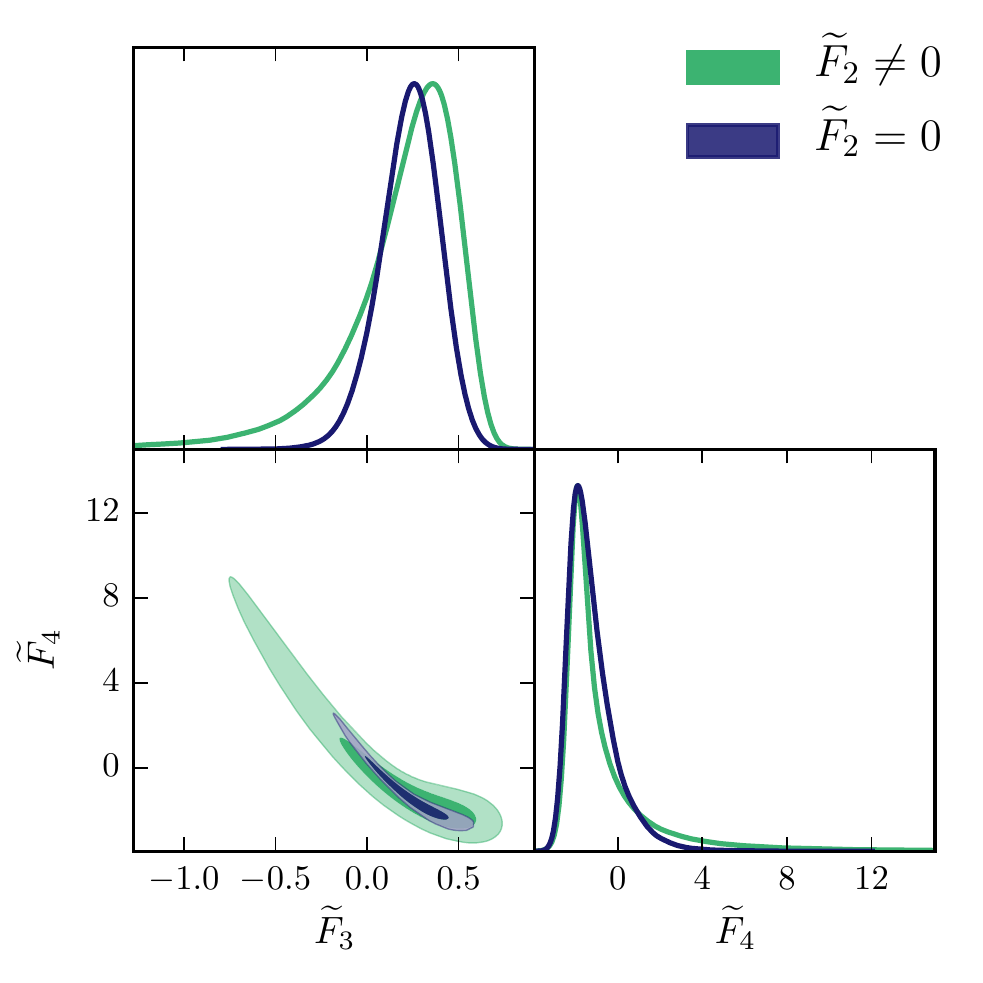}
\caption{Results for $\FT$ theories assuming $F_2 = 0$, compared with the general results, for $z<1$ (left) and all--$z$ (right). See Table~\ref{tab_results_F2_beta}. There are noticeable differences in the results for both $\tilde F_3$ and $\tilde F_4$, with the errors being underestimated in the $F_2 = 0$ case.}
\label{fig_F2}
\end{figure}

\begin{figure}[ht]
\centering
\includegraphics[width=.4\textwidth]{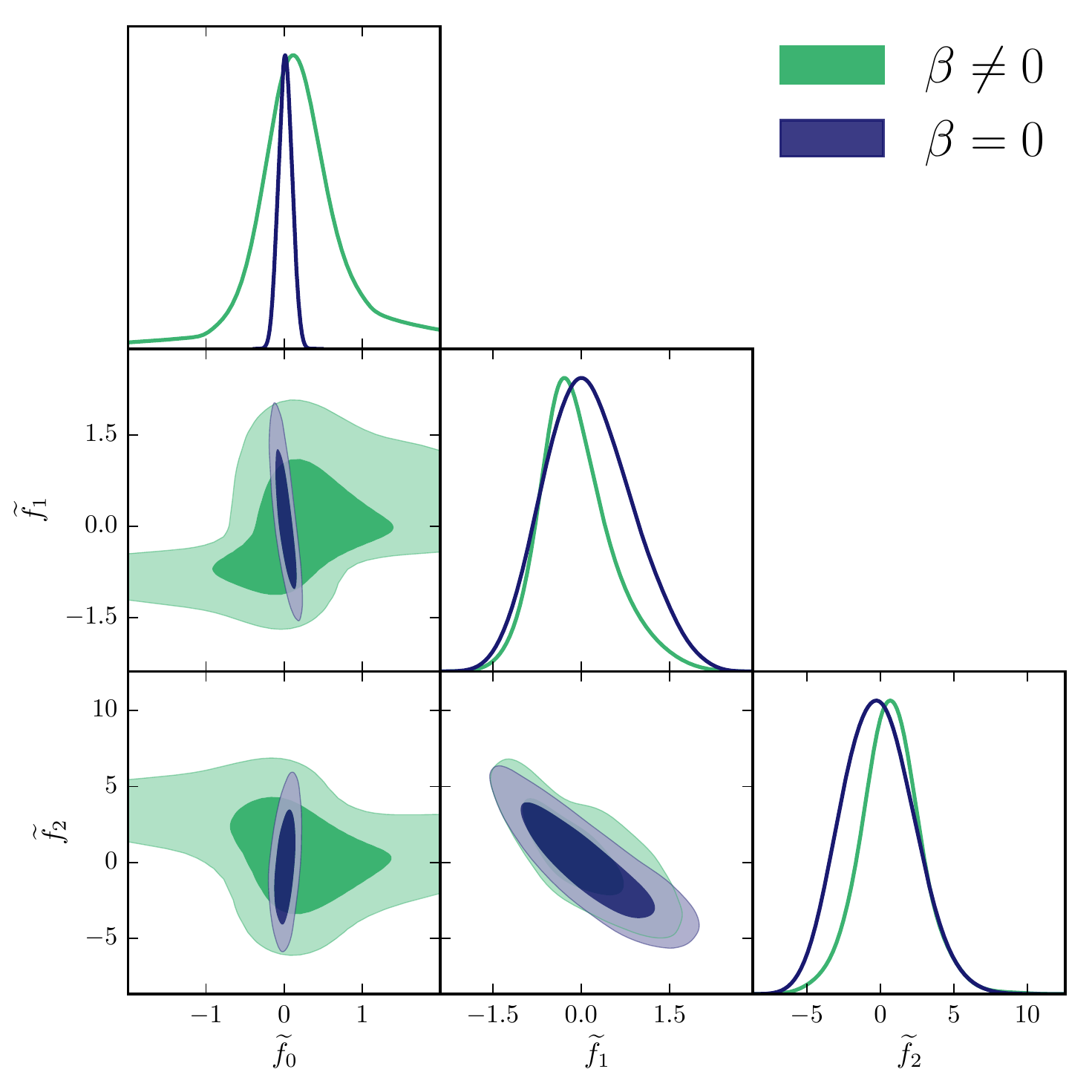}
\hspace{3em}
\includegraphics[width=.4\textwidth]{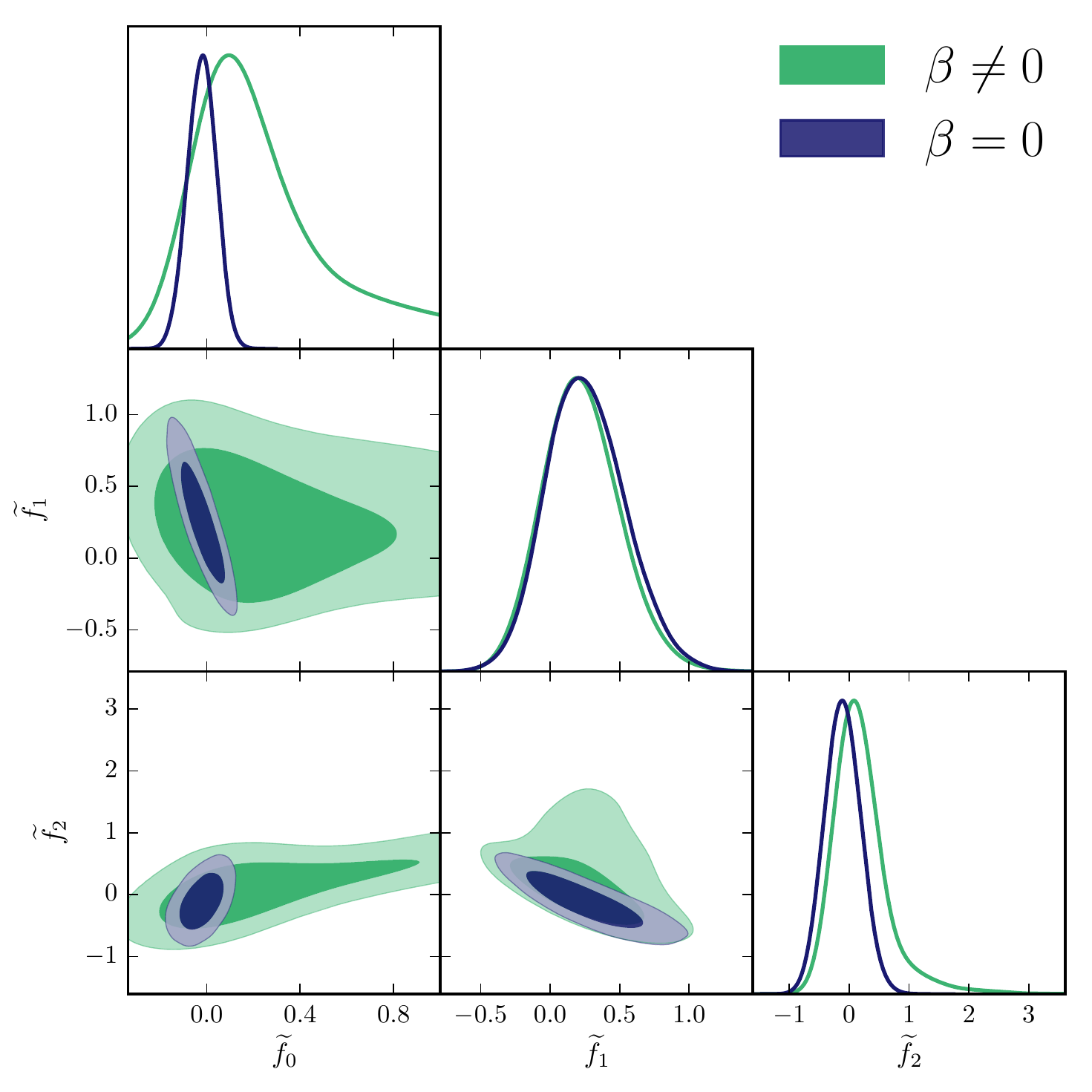}
\caption{Results for $f(R)$ theories assuming $(\alpha,\beta) = (1,0)$, for $z<1$ (left) and all--$z$ (right), compared with the general results. See Table~\ref{tab_results_F2_beta}. Imposing $\beta = 0$ has little effect on the bounds for $\tilde f_1$ and $\tilde f_2$, but the bounds $\tilde f_0$ are substantially modified.}
\label{fig_beta}
\end{figure}

The typical approach, commonly adopted in the literature when dealing with $f(R)$ and $\FT$ theories, is to assume that $\alpha = 1$ \cite{Aviles:2012ir} and $F_1 = 1$ \cite{bc} respectively. The consequence of these choices is that the cosmological value of Newton's constant is fixed to its Solar system value and naively extended to cosmological scales \cite{bc}. Moreover, another hypothesis is to assume that the current values of the second derivative of $f$ and $F$ are negligibly small at $R=R_0$ and $\mathcal T=\mathcal T_0$, i.e., $\beta$ and $F_2$ very close to zero respectively. This guarantees that the limiting cases of $f(R)$ and $\FT$ reduce to the $\Lambda$CDM model in intermediate redshift domains.
These choices on both the first and the second derivatives on the $f(R)$ and $F(\mathcal{T})$ gravitational Lagrangians today are very strict since they limit the extensions of GR to have a cosmological constant as limiting case. Hence, although reasonable,  these assumptions are not at all general. As a byproduct, the dark energy reconstruction is based on modeling the discrepancies with data by smoothing different functions under very tight assumptions and consequently the danger is to obtain \emph{model dependent} reconstructed cosmological models. Therefore, the \emph{inverse procedure} of obtaining potential candidates of $f(R)$ and $\FT$ from cosmography may be misleading.

The comparison between our approach and the usual approximations are shown in Figs.~\ref{fig_F2} and \ref{fig_beta} and Table~\ref{tab_results_F2_beta}. In particular, we find that:

\begin{itemize}

\item For $\FT$ gravity, setting $F_2=0$ has a very significant effect on both $\tilde F_3$ and $\tilde F_4$, whose 95\% errors are reduced by at least about a factor 3 (for $\tilde F_3$, all--$z$) up to 3 orders of magnitude ($\tilde F_4$, $z<1$).

\item
in the case of $f(R)$ theories, setting $\beta = 0$, as done in Ref. \cite{bc}, affects mainly the bounds on $\tilde f_0$, while $\tilde f_1$ and $\tilde f_2$ have essentially the same posterior distributions as in the $\beta \neq 0$ case. Errors of $\tilde f_0$ are about a factor 10 larger if we relax the assumption $\beta = 0$.
Results in \cite{cosmo9} provided a first insight about the intrinsic limitations of cosmography when dealing with $f(R)$
theories showing how naive priors on $f_R$ and $f_{RR}$ as those in Ref. \cite{Aviles:2012ir} were very limited. Ref. \cite{cosmo9} also showed that the cosmographic approach is unable to find the specific $f(R)$ model even when mock luminosity distance data are generated with a viable (Hu-Sawicki) $f(R)$ model cosmology. In other words, the cosmographic technique was proved unable to adequately constrain the $f(R)$ parameters responsible for a cosmological evolution when only mock luminosity distance data are used.

\end{itemize}

Thus, our results show unequivocally that the assumptions made in the literature when discussing constraints on $\FT$ and $f(R)$ theories from cosmography are in general wrong, and result in excessively strong bounds on the viability or parameter space of theories. In particular, we have shown that the usual assumption that $F_1 = \alpha = 1$ is completely superfluous, because both parameters can be eliminated with a suitable rescaling of the other parameters. Moreover, the assumptions $\tilde F_2 = \beta = 0$ strongly affect the posterior probabilities of the other model parameters, in such a way that error bars are grossly underestimated. Our analysis shows that a much larger portion of parameters space of these theories is actually allowed by cosmographic tests.

On the one hand, this is an indication of the general ability of these theories to mimic $\Lambda$CDM, especially at low redshift. On the other hand, it shows the limitations of the cosmographic approach in effectively constraining modified gravity theories.

As a complementary comment the results we have obtained in the previous sections show that the Gaussian process technique when applied to quintessence theories (see \cite{Nair:2013sna} and citations of that reference) remains more competitive than the cosmographic approach, even when other cosmological probes (BAO and $H(z)$) are combined with SNIa data.
Finally, let us mention that Refs. \cite{Aviles:2012ir, bc, cosmo9, delaCruz-Dombriz:2016rxm} only dealt with SNIa data whereas our present analysis has included both Union 2.1 supernovae catalogue, baryonic acoustic oscillation data and $H(z)$ differential age compilations, which indeed probe cosmology on different scales of the cosmological evolution. In this sense also, our analysis overcome previous attempts to prove the validity and competitiveness of the cosmographic approach.

\section{Conclusions}
\label{S4:Conclusions}

In this paper, we have extended previous cosmographic analyses to three classes of competing modified theories of gravity. In particular, we have considered quintessence, $\FT$ and $f(R)$ theories. We have tested these theories using cosmography without assuming any limitations on the parameter space for each of these gravitational theories. In doing so, we avoided the common treatment developed by the majority of previous authors, who were either unable to capture essential trends in the theories under consideration or underestimated the cosmographic bounds. In particular, by virtue of the fact that cosmography is a completely model-independent method based on the cosmological principle only, we derived constraints on the cosmographic series which do not depend essentially on any specific modifications of General Relativity. This allows one to put bounds on the cosmic coefficients of any theory without postulating the underlying model \emph{a priori}. We were therefore able to fit these theories using three different catalogues: Union 2.1 type Ia supernovae, BAO and $H(z)$ measures, with the support of the most recent Planck data. We employed the use of flat priors on all parameters with the exception of $r_{\rm s}(z_{\rm drag})$, in which a Gaussian prior was used, set at Planck's best values.

We first used data in the very small redshift regime $z<1$, using 22 and 551 data points for $H(z)$ and SNIa respectively. We then performed the same analyses but with all data points for every dataset, i.e., without limiting them to $z<1$. In this way, we were able to check the consistency of using finite Taylor expansions. 
This does not apply to BAO measurements, since all data points are confined inside $z<1$. Hence, the same data were used for both fits.

With these considerations in mind, we performed MCMC simulations using a Metropolis-Hastings algorithm and a Gelman-Rubin convergence diagnostics. We then performed the statistical analysis of the produced chains using publicly available \texttt{Python} codes.

As a standard way out of comparing our approaches with data, we first presented the $\Lambda$CDM results, in which the only free parameter turns out to be the total non-relativistic matter density parameter $\Omega_m$, as $\Omega_k$ is set to zero. We obtained excellent agreement between the $z<1$ and all--$z$ datasets, as slightly smaller $\Omega_m$ have been taken into account in our fits. We therefore investigated ways in which our constraints are affected by truncating the expansion of $H(z)$ at finite orders, by comparing results for the exact $\Lambda$CDM model and for the corresponding truncated series. We found the exact and truncated models agree within 1--$\sigma$ level, with an almost perfect overlap between the two cases for $z<1$.

The main results for every class of extended theories can be summarised as follows:
\begin{itemize}
\item For quintessence: all derivatives of the potential are compatible with zero inside the 2--$\sigma$ confidence levels, demonstrating that the Concordance model works perfectly well, while still allowing for some slight deviations, i.e., $\tilde V_0\in \mathcal O(0.1)$ and $\tilde V_1$ and $\tilde V_2$ constrained at the level of~$\mathcal O(1)$. The inclusion of the data at larger $z$ ($z>1$), as expected, results in a reduction of the errors on parameters, particularly for higher derivative terms.

\item For $\FT$ theories we again found that all model parameters are compatible with zero at about the 1--$\sigma$ level, while the 2--$\sigma$ level enables for quite larger parameter ranges, especially for higher derivatives, i.e., $\tilde F_3$, $\tilde F_4$. In fact, $\tilde F_3$ and $\tilde F_4$ are only constrained at the level of $\mathcal O(10^2)$ and $\mathcal O(10^2)$ levels for $z<1$ data, and $\mathcal O(1)$ and $\mathcal(10)$ for the full dataset. The best-fit points are still relatively close to $\tilde F_i = 0$, but the posterior probabilities are far from Gaussian, with long tails which extend to either positive or negative values.

\item For $f(R)$ theories, we found that all model parameters were compatible with zero at about the 1--$\sigma$ level, albeit the shapes of contour plots being different from pure ellipses. Indeed, such shape also changes as one moves from the $z<1$ analysis to the all--$z$ analysis.

\end{itemize}

In all the theories considered, we have also compared our statistical results to the widely used AIC and BIC criteria. We found that, expanding the standard cosmological model till the third order, the corresponding third orders of any modified theories studied here seem to be statistically favoured by using the AIC criteria, however disfavoured in the case of BIC.

Finally we compared our treatment, which was solely based on
the most general assumptions, to previous approaches, noticing a discrepancy in the results. Thus we showed
the undesirable model-dependence which existed in the previous literature for several classes of reconstructed extended dark-energy theories.

Future work will take higher-redshift catalogues into account in order to fully characterise the cosmographic approach established here, with different windows of data points. This will be helpful in determining the effective dark energy evolution, as obtained from several classes of modified theories of gravity, in different cosmological eras.

%
\acknowledgments

A.d.l.C.D. acknowledges financial support from the University of Cape Town (UCT) Launching Grants
programme, National Research Foundation (NRF) grant 99077 2016-2018, ref.~No.~CSUR150628121624, MINECO (Spain) projects FIS2014-52837-P, FPA2014-53375-C2-1-P, Consolider-Ingenio MULTIDARK CSD2009-00064 and CSIC I-LINK1019 and would also like to thank the Instittuto de F\'isica Te\'orica (IFT UAM-CSIC, Madrid Spain) for its support via the KA107 action of the Erasmus$+$ Call for international Mobility and the EU COST Action CA15117. P. K. S. D. and L. R. thank the National Research Foundation (NRF) for financial support.

\bibliography{SINGLEFILE_Biblio}

\end{document}